\documentclass[10pt,trans]{IEEEtran}

\usepackage{stfloats}
\usepackage{amsfonts}
\usepackage{amssymb}
\usepackage{amsthm}
\usepackage{cite}
\usepackage[cmex10]{amsmath}

\usepackage{float}
\usepackage{color}
\usepackage{algorithm}
\usepackage{algorithmic}
\usepackage{multirow}
\usepackage[normalem]{ulem}
\usepackage{tabularx}
\usepackage{subfigure}
\usepackage{fancybox,dashbox}
\usepackage{bbm}
\usepackage[dvipsnames]{xcolor}

\newtheorem{theorem}{Theorem}

\newtheorem{proposition}{Proposition} 

\newtheorem{definition}{Definition}
\newtheorem{remark}{Remark}

\ifCLASSINFOpdf
\usepackage[pdftex]{graphicx}
\DeclareGraphicsExtensions{.pdf,.jpeg,.png}
\else
\usepackage[dvips]{graphicx}
\DeclareGraphicsExtensions{.eps}
\fi

\usepackage{hyperref}
\usepackage[hyphenbreaks]{breakurl}
\usepackage{geometry}
\begin{document}
	
	\title{Task Offloading for Large-Scale Asynchronous Mobile Edge Computing: An Index Policy Approach}
	\author{Yizhen Xu, Peng Cheng, Zhuo Chen, Ming Ding,  Branka Vucetic, and Yonghui Li
		\thanks{Y.~Xu, B.~Vucetic, and  Y.~Li are with the School of Electrical and Information Engineering, the University of Sydney, Australia, (e-mail: yizhen.xu@sydney.edu.au; branka.vucetic@sydney.edu.au; yonghui.li@sydney.edu.au). P. Cheng is with the Department of Computer Science and Information Technology, La Trobe University, Melbourne, VIC 3086, Australia, and also with the School of Electrical and Information Engineering, the University of Sydney, Sydney, NSW 2006, Australia (e-mail: p.cheng@latrobe.edu.au; peng.cheng@sydney.edu.au). Z.~Chen and D.~Ming are with CSIRO DATA61, Australia (e-mail:ming.ding@data61.csiro.au; zhuo.chen@ieee.org). The work of P. Cheng was supported by ARC under Grant DE190100162 and DP210103410. The work of Y. Li was supported by ARC under Grant DP190101988 and DP210103410. The preliminary results were presented in \cite{9054413}. \textit{(Corresponding Authors: Peng Cheng; Yonghui Li)}}}

	\maketitle
\markboth{Accepted by IEEE Transactions on Signal Processing (Full Version)}{}	
\begin{abstract}

Mobile-edge computing (MEC) offloads computational tasks from wireless devices to network edge, and enables real-time information transmission and computing. Most existing work concerns a small-scale synchronous MEC system. In this paper, we focus on a large-scale asynchronous MEC system with random task arrivals, distinct workloads, and diverse deadlines. We formulate the offloading policy design as a restless multi-armed bandit (RMAB) to maximize the total discounted reward over the time horizon. However, the formulated RMAB is related to a PSPACE-hard sequential decision-making problem, which is intractable. 
To address this issue, by exploiting the Whittle index (WI) theory, we rigorously establish the WI indexability and derive a scalable closed-form solution. Consequently, in our WI policy, each user only needs to calculate its WI and report it to the BS, and the users with the highest indices are selected for task offloading. 
Furthermore, when the task completion ratio becomes the focus, the shorter slack time less remaining workload (STLW) priority rule is introduced into the WI policy for performance improvement.
When the knowledge of user offloading energy consumption is not available prior to the offloading, we develop Bayesian learning-enabled WI policies, including maximum likelihood estimation, Bayesian learning with conjugate prior, and prior-swapping techniques. 
Simulation results show that the proposed policies significantly outperform the other existing policies. 

\end{abstract}
	
	
\section{Introduction}

\subsection{Mobile Edge Computing}
\label{MEC}
The exponential growth in smart device adoption is accelerating the ubiquitous Internet of Things (IoT) \cite{7488250, 7786106}, and catalyzing the development of computation-intensive applications, including augmented reality (AR), face recognition, interactive online gaming, and autonomous driving. 
However, a computationally tedious task is unlikely to be executed at the local mobile device due to its resource constraint. Instead, such tasks can be offloaded to the remote cloud with abundant computation, storage, and energy resources \cite{5445167, zhang2010cloud, 6897914}. Despite the computational efficiency, the long communication distance between smart devices and the remote cloud inevitably introduces a high transmission latency, resulting in an unsatisfactory user quality of experience (QoE), especially for numerous real-time delay-sensitive applications.   
	
To enable computationally intensive applications with sensitive delay requirements, mobile edge computing (MEC) \cite{hu2015mobile, 7901477, 7883826} has recently emerged as a promising paradigm. Compared with its cloud counterpart, MEC pushes the computing and storage capability to the network edge that is much closer to devices. As a result, 
a device can offload its tasks to a proximal MEC server at a base station (BS) or access point (AP), and then collect the subsequent results from the MEC server. This generates the benefits of low latency and reduced mobile device energy consumption. Essentially, task offloading involves joint radio-and-computation resource allocation among multiple users. 
    
A variety of specific task offloading policies have been designed in recent years. Most of them are focused on small-scale synchronous MEC systems, where tasks for a limited number of different users arrive simultaneously. 
In this context, the computational resources are relatively abundant, and the policy design was usually formulated as a static centralized optimization problem, where the energy consumption and latency serve as the principal performance indicators \cite{7227025, 8327930, 8611399}. 
For example, to minimize the weighted sum of energy consumption and end-to-end delay, the authors in~\cite{7227025} formulated the task offloading problem into a non-convex quadratically constrained quadratic program (QCQP), and semi-definite relaxation and randomization mapping based algorithms were proposed to achieve a near-optimal offloading performance. 
 A joint radio and computational resource allocation scheme was investigated in \cite{8327930}, and the offloading was formulated as a mixed-integer nonlinear programming problem (MINLP). A game theory-based approach was proposed to handle this problem. Recently, the offloading problem was formulated in~\cite{8611399} as a nonconvex optimization one, which aimed to maximize the weighted sum computation efficiency by imposing the constraints on local computation capability and energy resources. 
	
\subsection{Task Offloading for Large-Scale Asynchronous MEC}
A wide range of emerging massive machine type communication applications, such as industrial automation, and smart transportation~\cite{6774858}, introduce the concept of large-scale asynchronous MEC systems. Typically, these applications involve a variety of tasks each with its own execution deadline. 
Furthermore, task arrival patterns for a massive number of users exhibit notable stochasticity, featured by random and asynchronous task arrivals, distinct workloads, and diverse deadlines (see Fig.~\ref{fig: system state}). Due to the deadline constraints on tasks and limited computational resources (with respect to a massive number of users), task offloading policy design for large-scale asynchronous MEC systems becomes extremely challenging. 
Task offloading policy should focus on the user selection and be implemented on a dynamic and real-time basis, considering both the task criticality and energy consumption. 
To the best of our knowledge, the offloading policy design in a large-scale asynchronous MEC system is still an open challenge.

\textcolor{black}{
In this paper, we aim to address this challenge and propose an index-based task offloading policy. To adapt to the dynamic nature of task arrivals, we deviate from the classical static centralized optimization techniques with hard constraints, and turn to the bandit theory to capture the stochastic behavior in tasks. 
We formulate the offloading policy design as a restless multi-armed bandit (RMAB) problem. Mathematically, MAB is a sequential decision model with a set of arms to choose from for the total reward maximization~\cite{5398950, 6035799, whittle1988restless}. At each round, only a subset of arms can be selected and their states will change, while the others remain frozen. 
Removing such restrictions in the MAB, the RMAB allows the states of all arms to evolve over time regardless of the actions.
In our setting, we treat each user as an independent restless arm, and the arm state is represented by user task criticality including the remaining number of subtasks and remaining time to deadline. 
We then design a reward function to strike a promising balance between two conflicting goals, i.e., minimizing the energy consumption and maximizing the task completion ratio. In this case, ``playing" an arm at each time slot is equivalent to selecting a user to offload its tasks. }
	
Our goal is to maximize the total discounted reward over the time horizon for the formulated RMAB, resulting in a new task offloading policy. However, the RMAB is generally PSPACE-hard and intractable~\cite{papadimitriou1999complexity}. To address this issue, we develop a novel method based on Whittle index (WI)~\cite{whittle1988restless}, so that multiple arms can be decoupled and the original $N$-dimensional problem reduces to $N$ independent 
$1$-dimensional ones. 
The key advantage of our WI offloading policy lies in its excellent scalability and low computational complexity, enabling fast user selection in task offloading. 
At each time slot, each user only needs to separately calculate its scalar WI in closed form which provides a proxy to measure its task criticality. Each user then reports its WI to the BS, and the users with the highest indices are selected for task offloading. Besides, the WI policy can be implemented in a totally distributed manner.
	
Specifically, we first consider the scenario where the perfect knowledge of the user offloading energy consumption is available at users. 
We exploit the WI theory and rigorously establish the indexability of the RMAB through the inductive method, which theoretically guarantees the existence of WI for our RMAB. On this basis, we derive a closed-form expression in terms of the task state and energy consumption for the WI computation. 
Furthermore, when the task completion ratio becomes the focus, the shorter slack time less remaining workload (STLW) priority rule is introduced into the WI offloading policy for performance improvement, referred to as STLW-WI policy. 
On the other hand, when the knowledge of user offloading energy consumption is not available prior to the offloading, the WI policy can not be directly applicable. To address this challenge, we develop Bayesian learning-enabled WI policies. In specific, we first integrate the WI policy with the maximum likelihood estimation (MLE) technique. Then, to further improve the performance, we propose a novel Bayesian learning with WI policy (BL-WI) given the conjugate prior. Finally, a refinement mechanism (PSBL-WI) based on prior-swapping is proposed for a fast inference given the non-conjugate prior.   
It is verified by simulation that the proposed WI policy can achieve much better performance in terms of the total discounted reward, compared with several existing offloading policies. For the completion ratio-oriented task offloading, our STLW-WI policy achieves a higher task completion ratio. 
When the user offloading energy consumption is unknown, our Bayesian learning-enabled WI policy can also achieve a favorable performance compared to the original WI policy. 

\color{black}

\subsection{Contribution}
The main contributions of this paper can be summarized as follows.
\begin{itemize}
\item We develop an RMAB framework to enable task offloading for large-scale asynchronous MEC in a realistic setting. We propose a novel WI offloading policy through establishing the WI indexability and deriving a scalable solution with closed-form expression. When the knowledge of user offloading energy consumption is unknown prior to offloading, novel MLE-WI, BL-WI and PSBL-WI offloading policies from the Bayesian learning perspective are developed.

\item The developed WI method offers a potential low-complexity solution to a series of communication/computation resource scheduling problems (e.g.,  scheduling in a power-aware server farm \cite{5703092}), which typically involves complicated combinatorial optimizations. The developed WI method also addresses a challenging general RMAB problem in a dynamic environment with heterogeneous rewards and non-identical transition probabilities. 
\end{itemize}

\subsection{Related Work}
In addition to the work introduced in Section \ref{MEC}, task offloading policy for MEC has been extensively studied in the literature. Two comprehensive surveys on various task offloading policy designs were provided in \cite{8016573, 7879258}. The work in \cite{8638800} designed a new MEC system to satisfy the ultra-reliable low-latency requirements in mission-critical applications. Specifically, a two-timescale association between user and server was proposed by utilizing the Lyapunov optimization and matching theory. For both time division multiple access (TDMA) and orthogonal frequency division multiple access (OFDMA), corresponding offloading policies have been developed in \cite{7762913}, aiming to minimize the user weighted sum energy consumption given the constraints of the user average latency. 

As a special case of reinforcement learning \cite{sutton2011reinforcement}, the stateless MAB techniques have been applied to MEC systems \cite{8058414, 8627987, 8790775, 9082866, 8762108, 8887222}. In stateless MAB, the arms do not have any specific state. Each arm, when played, offers an i.i.d random reward drawn from a distribution with an unknown mean. 
The authors in \cite{8058414} developed an energy-aware mobility management scheme based on MAB to perform MEC selection. In \cite{8627987}, an adaptive learning task offloading policy was proposed for vehicle edge computing based on the MAB theory. 
In \cite{8790775}, the authors considered an edge service replacement problem, where they applied contextual combinatorial MAB to estimate users' demand based on side information. 
An MAB online learning algorithm referred to as utility-table learning was proposed in \cite{9082866} to determine the optimal workload balance among MEC servers. In \cite{8762108}, the authors proposed an online task offloading policy based on the non-stationary MAB model, aiming to minimize the long-term total costs including latency, energy consumption and switching cost. Under the MAB framework, a two-stage resource sharing and task offloading strategy were developed in \cite{8887222}. 
By contrast, the RMAB in this paper can be categorized into the stateful bandit model \cite{7498076}, where every arm is associated with some finite state space and the state evolves as a Markov process. When an arm is selected, the reward is drawn from some stationary distributions based on the current arm state.  

There are several other works leveraging the WI theory first established in \cite{whittle1988restless}. According to \cite{gittins2011multi, ayesta2019unifying}, the WI solution in \cite{whittle1988restless} does not hold in general, and there is no unified solution that can cover all the RMAB problems. Consequently, the establishment of indexability needs to be studied for the individual problem. 
For example, the problem formulation in \cite{5605371} suits a restless Bernoulli bandit with a two-state Markov chain, and the establishment of indexability highly depends on the transition probability of wireless channel occupancy. In contrast, the formulation in \cite{8295041} suits an RMAB problem with a static reward and identical state transition probability for each arm. In this paper, we aim to solve a new RMAB problem with heterogeneous rewards and a non-identical transition probability for each arm.

\subsection{Organization}
The rest of the paper is organized as follows. In Sections \ref{section2} and \ref{sec: RMAB Formulation}, we discuss the MEC system model and formulate the offloading problem as an RMAB, respectively. 
In Section \ref{sec: WI}, we establish the indexability of the RMAB and develop a WI offloading policy. A Bayesian learning enabled WI offloading policy is proposed in Section \ref{sec: learning}. Simulation results are presented in Section \ref{sec: Numerical Results} followed by conclusions in Section \ref{section 7}.

\color{black}
\textit{Notation}: $\mathcal{N}(\mu,\Sigma)$ denotes the Gaussian distribution with a mean $\mu$ and a variance $\Sigma$. $\Gamma(\cdot)^{-1}$ denotes the Gamma inverse function. $x^{+} = \max(x,0)$. $\mathbbm{1}(\cdot)$ is the indicator function. $\binom{N}{M}$ denotes the combinations of selecting distinct $M$ items out of $N$. 
For convenience, we also list most important symbols in Table \uppercase\expandafter{\romannumeral1}. 
\begin{table*}[t]
	\renewcommand\arraystretch{2}
	\centering
	\caption{TABLE OF SYMBOLS}
	\begin{tabular}{|c|c|}
		\hline
		Variable & Description \\
		\hline
		$\beta$ & the discount factor \\
		\hline
		$B_{i,j}$ & the total number of subtasks for the $i$-th user in the $j$-th task \\
		\hline
		$b_{i,t}$ & the number of unfinished tasks in the $i$-th user at the $t$-th time slot \\
		\hline
		$C_i$ & the number of CPU cycles required to process $1$ bit data by the $i$-th user \\
		\hline 
		$\epsilon_i$ & $\epsilon_i \sim \mathcal{N}(0, \Sigma_i)$; the measurement noise with noise variance $\Sigma_i$ \\
		\hline
		$E_{i}^{\text{loc}}$ & the local computing energy consumption by the $i$-th user \\
		\hline 
		$E_{i, j}^{\text{off}}$ & the offloading energy consumption by the $i$-th user during the $j$-th task \\
		\hline 
		$E_{i, j}^{\text{sav}}$ & $E_{i, j}^{\text{sav}} = k_i E_{i}^{\text{loc}} - E_{i, j}^{\text{off}}$; the energy consumption saving from the offloading for $k_i$ subtasks \\
		\hline
		$e_{i,j}^{\text{sav}}$ & $e_{i,j}^{\text{sav}} = E_{i,j}^{\text{sav}} + \epsilon$; a noisy observation version of the actual energy saving $E_{i,j}^{\text{sav}}$ \\
		\hline
		$\gamma_{i,j}$ & the number of observations for the $i$-th user during $j$-th task \\
		\hline
		$h_{i,j}$ & the channel gain of the $i$-th user during the $j$-th task \\
		\hline
		$k_i$ & the number of subtasks which can be processed  by the MEC server in each time slot \\
		\hline
		$\kappa_{i,j}$ & small-scale fading channel power gain for the $i$-th user during $j$-th task \\
		\hline
		$l_{i,t}$ & $l_{i,t} \triangleq \tau_{i,t} - b_{i,t} / k_i$; the slack time of the $i$-th user at the $t$-th time slot \\
		\hline
		$M$ & the number of MEC servers in the system \\
		\hline
		$N$ & the number of users in the system \\
		\hline
		$r_{i,j}$ & the achievable transmission rate of the $i$-th user during $j$-th task \\
		\hline
		$U_i$ & the CPU frequency of the $i$-th user \\
		\hline
		$s_{i,t}$ & $s_{i,t} = \left(\tau_{i,t}, b_{i,t} \right)$; the state of the $i$-th user at the $t$-th time slot \\
		\hline
		$\textbf{S}_t$ & $\textbf{S}_t = \left(s_{1,t}, \cdots, s_{N, t}\right)$; the system state at the $t$-th time slot \\
		\hline
		$t_{i,j}^{d}$ & the deadline of the $i$-th user's $j$-th task \\
		\hline
		$\tau_{i,t}$ & $\tau_{i,t} \triangleq t_{i,j}^{d} - t + 1$; the number of remaining time slots to $t_{i,j}^{d}$ \\
		\hline
		$u_{i,t}$ & $u_{i,t} \in \left\{0, 1\right\}$; the offloading action on the $i$-th user at the $t$-th time slot \\
		\hline
		$\textbf{u}_t$ & $\textbf{u}_t = \left(u_{1,t}, \cdots, u_{N, t}\right)$; the actions taken by the BS for each user at the $t$-th time slot \\
		\hline
		$\omega_i(s_{i,t})$ & the WI of the $i$-th user given its current state $s_{i,t}$ \\
		\hline
		$X_{i,j}(t)$ & $X_{i,j}(t) = \left\{e_{i,j,1}^{\text{sav}}, \cdots, e_{i,j,\gamma_{i,j}}^{\text{sav}}\right\}$ the $i$-th user's energy saving observation set up to the $t$-th time slot \\
		\hline
	\end{tabular}\\
	\label{tab:1}
\end{table*}
		
\color{black}
	
\section{System Model}
\label{section2}
\color{black}
\subsection{Large-Scale Asynchronous Task Arrival Model}
		
\begin{figure}[t]
\centering
{\includegraphics[width=0.5\textwidth]{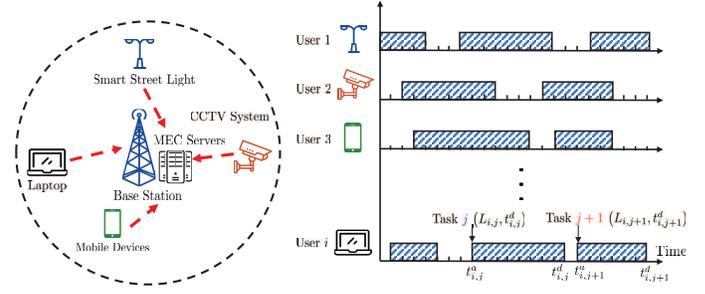}}
\caption{A large-scale asynchronous MEC system (left) with an illustrative asynchronous task arrival pattern (right). The area of the shaded rectangular indicates the size of the task $B_{i,j}$.}
\label{fig: system state}
\end{figure}

\textcolor{black}{Consider a large-scale asynchronous MEC system consisting of a BS and $N$ static users (indexed by $i \in \left\{1, \cdots, N\right\}$) shown in Fig. \ref{fig: system state}, where $N$ is reasonably large. 
The system operates in a time-slotted structure, indexed by $t$. The BS is equipped with $M (M < N)$ independent MEC servers, and we assume that each MEC server can serve at most one user at each time slot. 
Each user is running computation-intensive and delay-sensitive tasks with stochastic arrival patterns to be elaborated later. Each task is relatively large and can be further partitioned into a number of subtasks to be processed sequentially \cite{8663994}. It is assumed that the local computation capability of each user is not powerful enough to complete a task on time. Therefore, a user seeks assistance from the MEC server by offloading some subtasks for faster execution. As we only have $M$ MEC servers (limited computational resources), at most $M$ users can be selected to perform task offloading at each time slot. The number of possible combinations is $\binom{N}{M}$, which is usually extremely huge to handle\footnote{For example, when $M=30$, $N=100$, we have $\binom{N}{M} \approx 3 \times 10^{25}$.}.}
    

Next, we specify the asynchronous task arrival pattern with an example shown in Fig.~\ref{fig: system state} (a more detailed example is shown in Fig.~\ref{fig: offloading result}). The task arrival time and its deadline vary for each user. For the $i$-th user, at the $t_{i,j}^{a}$-th time slot, a new task $j$ arrives and reveals the number of subtasks $B_{i,j}$ and the task deadline $t_{i,j}^{d}$. 
Without loss of generality, a user's subtasks are assumed to have equal size ($l_i$ bits). Upon arrival, the task starts to be processed and will be removed from the user buffer at the end of the $t_{i,j}^{d}$-th time slot. It is worth noting that $t_{i,j}^{a}$, $t_{i,j}^{d}$, and $B_{i,j}$ are discrete random variables. At the beginning of the $t$-th time slot, if the $i$-th user is idle, a new task will arrive with the probability $Q_i$.

\subsection{Computation Model}
Each user is assumed to be able to locally process one subtask in each time slot. For the $i$-th user, the number of CPU cycles required to process $1$ bit data is denoted by $C_i$, which may be different for various users~\cite{miettinen2010energy}. We denote the CPU frequency of the $i$-th user by $U_i$, then the local execution cycles for one subtask with $l_i$ bits can be calculated by $C_{\text{loc}} = C_i l_i / U_i$. Since the user is usually operating at a constant $U_i$ for the sake of energy efficiency~\cite{burd1996processor, 8606442}, the computing power for each CPU cycle can be calculated as $P_{0,i} = \lambda U_i^2$, where $\lambda$ is a power coefficient depending on the chip architecture~\cite{8606442}. 
In this case, the local computing energy consumption $E_{\text{loc}, i}$ for the $i$-th user to process one subtask can be calculated as   
\begin{equation}
    E_{i}^{\text{loc}} =  \lambda U_i C_i l_i.
\end{equation}
	
For MEC servers, we assume that their CPU frequency is a constant $U_{s}$. Accordingly, $k_i = \left \lfloor \frac{U_s}{U_i} \right \rfloor$ is the number of subtasks which can be processed in each time slot by an MEC server. That is, if the $i$-th user is selected to perform task offloading, it can transmit $k_{i}$ subtasks to the server.

\subsection{Communication Model}
Following~\cite{7762913}, the offloading process at each time slot can be divided into three steps: 1) a selected user uploads some subtasks to the MEC server; 2) the MEC server processes the subtasks; 3) the results are transmitted back to the user.  
As the results are usually of small size~\cite{7563449}, the downloading time and associated energy consumption can be omitted. Therefore, in the whole process, the user offloading energy consumption mainly comes from the uploading phase. 

As at most $M$ users can be selected at each time slot, we adopt OFDMA scheme for data transmission. For the $i$-th user during the $j$-th task, the achievable transmission rate can be calculated as
\begin{equation}
    r_{i,j} = W_i \log_2\left(1+\frac{P_{i}^{tx} 
    h_{i,j}}{N_0 W_i}\right),
\end{equation}
where $W_i$ is the bandwidth, $P_{i}^{tx}$ is the transmission power, $N_0 W_i$ is the noise power. In addition, $h_{i,j} = \kappa_{i,j} g_0 (d_0/d_{i,j})^{\iota}$ denotes the channel gain of the $i$-th user~\cite{7956189} with $\kappa_{i,j}$ being the small-scale fading channel power gain. 
In this paper, we adopt a widely used block fading channel model~\cite{ 8611399, 7541539}, and $\kappa_{i,j}$ keeps constant during the $j$-th task but varies independently from task to task, where $g_0$ is the path-loss constant, $\iota$ is the path-loss exponent, $d_0$ is the reference distance, and $d_{i,j}$ is the transmission distance between the $i$-th user to the BS. 
Therefore, the required transmission time $t$ for sending $k_i$ subtasks can be calculated as $t_i = (k_i) l_i / r_{i,j} $. We assume that the length of a time slot is relatively large, so that it is always larger than the transmission time $t_i$. As a result, the offloading energy consumption of the $i$-th user during the $j$-th task is calculated as 
\begin{equation}
    E_{i, j}^{\text{off}} = t_i P_{i, j}^{tx} = \frac{k_i l_i}{r_{i,j}} P_{i, j}^{tx}.
\end{equation}
\color{black}

We need to consider both the task completion ratio and user's energy consumption under the limited computational resource ($M < N$).  
At this point, our goal is to strike a promising balance between energy consumption and task completion ratio. Furthermore, the large-scale stochastic task arrivals require the offloading policy to perform user selection dynamically at each time slot with a low computational complexity, which is quite challenging.

\section{Restless Multi-armed Bandit Formulation} 
\label{sec: RMAB Formulation}

To address the above challenges, we develop a novel WI policy that enables fast user selection at each time slot, in which offloading priorities are indicated by the value of WI.
For the specific implementation, we first formulate the offloading policy design as an RMAB \cite{whittle1988restless} to capture the randomness in tasks arrivals, workload and their deadlines. Essentially, we treat the remaining number of subtasks and remaining time to deadline of each task as the state of an arm. 
The restless nature of the state naturally follows, because the state will also change even though the user is not selected for offloading. Under this formulation, we also create a new reward function as the performance metric, taking into account both deadline requirements and user offloading energy consumption. 

In the following subsections, we elaborate on the RMAB with five key factors, including the system state, action, state transition, reward function, and objective function. 

\subsubsection{System State}
For the $i$-th user, its state $s_{i,t}$ is represented by its current $j$-th task state at the beginning of the $t$-th time slot, i.e., $s_{i,t} = (\tau_{i,t}, b_{i,t})$, where $\tau_{i,t} \triangleq t_{i,j}^{d} - t + 1$ is the remaining time slots to the task deadline $t_{i,j}^{d}$, and $b_{i,t}$ is the number of the unfinished subtasks. 
If there is no task, then $s_{i,t} = (0,0)$. Accordingly, $s_{i,t}$ can be written in a compact form as
\begin{equation}
s_{i,t} =
\begin{cases}
(0,0), & \mbox{no task}; \\
(\tau_{i,t}, b_{i,t}), & \mbox{otherwise}.
\end{cases}
\label{eq:state}
\end{equation}
Collecting the states of $N$ users, the system state ${\bf{S}}_t$ at the $t$-th time slot is denoted by ${\bf{S}}_t \triangleq \left(s_{1,t}, \cdots, s_{N,t}\right)$.

\subsubsection{Action}
At the beginning of each time slot, the action taken by the BS determines $M$ users (among $N$) which could offload their subtasks to the MEC. We define the action as ${\bf{u}}_t = \left(u_{1,t}, \ldots, u_{N,t}\right)$, where $u_{i,t}\in \left\{0,1\right\}$. %
When $u_{i,t} = 0$, task offloading is not allowed. When $u_{i,t} = 1$, the user will be selected to perform task offloading. 
	
\subsubsection{State Transition} 
As mentioned before, an MEC server can process at most $k_{i}$ subtasks compared to one subtask processed locally at the $i$-th user each time slot. Therefore, if a user can perform task offloading, the remaining number of subtasks will be reduced by $k_{i}$ maximally. 
If the user is idle at the $t$-th time slot, i.e., $s_{i,t}=(0,0)$ , a new task will arrive with probability $Q_i$ at the $(t+1)$-th time slot. 
Given the current state $s_{i,t}$ and the action $u_{i,t}$, the next state $s_{i,t+1}$ can be expressed by 

\begin{itemize}
\item If $\tau_{i,t} \ge 2$, 
    \begin{equation}
    s_{i,t+1}=
    \begin{cases}
    \left(\tau_{i,t}-1,(b_{i,t} - k_i)^{+}\right), \hspace{0.5cm} \mbox{if } u_{i,t} = 1; \\
    \left(\tau_{i,t}-1,(b_{i,t} - 1)^{+}\right), \hspace{0.65cm} \mbox{if }  u_{i,t} = 0; \\
    \end{cases}
    \end{equation}
\item If $\tau_{i,t} = 1$, 
    \begin{equation}
    s_{i,t+1}=
    \begin{cases}
    \left(t_{i, j+1}^{d} - t, B_{i, j+1}\right), & \mbox{with Prob. } Q_i \\
    (0,0), & \mbox{with Prob. } 1-Q_i;
    \end{cases}
    \end{equation}
\item If $s_{i,t} = (0,0)$, (assuming the index of the last task is $j$),
    \begin{equation}
    s_{i,t+1}=
    \begin{cases}
    \left(t_{i, j+1}^{d} - t, B_{i, j+1}\right), & \mbox{with Prob. } Q_i \\
    (0,0), & \mbox{with Prob. } 1-Q_i;
    \end{cases}
    \end{equation}
\end{itemize}
where $x^{+} = \max(x,0)$. Note that when $\tau_{i,t} = 1$, the task of the $i$-th user will reach its deadline and be removed from the user at the end of the $t$-th time slot. Then, at the beginning of the $(t+1)$-th time slot, the $j+1$-th task with $B_{i, j+1}$ subtasks and deadline $t_{i,j+1}^{d}$ will arrive with probability $Q_i$. 
	
\subsubsection{Reward Function} 
Here, we create a reward function in \eqref{eq: Reward} to balance the user offloading energy consumption and deadline requirements, 
\begin{figure*}[t]
	\begin{equation}
	\begin{aligned}
	& R(s_{i,t}, u_{i,t}) \\
	& =
	\begin{cases}
	E_{i,j}^{\text{sav}}u_{i,t} , &\mbox{if }  \tau_{i,t} > 1, b_{i,t} > 0; 
	\\
	E_{i,j}^{\text{sav}} u_{i,t}  -  F\left( \left[b_{i,t}-k_{i}u_{i,t} - (1-u_{i,t})\right]^{+}\right), 
	&\mbox{if } \tau_{i,t} = 1, b_{i,t} > 0; 	
	\\
	0, \hspace{5.95cm} 	&\mbox{otherwise.}
	\end{cases}
	\label{eq: Reward}
	\end{aligned}
	\end{equation}
	\hrulefill
\end{figure*}
where $E_{i,j}^{\text{sav}} = \left(k_i E_{i}^{\text{loc}} - E_{i,j}^{\text{off}} \right)$ is the energy consumption saving from the offloading for the $k_i$ subtasks. 
The penalty function is denoted by $F(x) = \alpha x^2 $ with $x$ indicating the number of unfinished subtasks, and $\alpha$ is the penalty parameter used to adjust penalty for unfinished tasks. \textcolor{black}{\footnote{The penalty function is widely used in Markov decision problem settings with specific form varying from case to case.}}
The key points can be highlighted as follows. 
\begin{itemize}
	\item When $\tau_{i,t} > 1$ and $b_{i,t} > 0$, the task has not reached its deadline, the reward is related to energy consumption saving $E_{i,j}^{\text{sav}}$ if performing task offloading ($u_{i,t}=1$).  
	\item When $\tau_{i,t} = 1$, the task will be removed at the end of the $t$-th time slot. If the task cannot be completed by its deadline, a penalty measured by the number of unfinished subtasks is imposed.
	\item The benefit of the reward function is to strike a balance between the energy consumption and deadline requirements. For example, putting a priority on the deadline leads to a larger $\alpha$. By contrast, reducing $\alpha$ can increase energy savings for the battery-powered IoT devices to prolong their lifetime. 
\end{itemize}

\subsubsection{Objective}
Our objective is to find a policy $\mathcal{G}$ to maximize the expected total discounted system reward with the constraint of the limited computational resources, which is defined by

\begin{equation*}
\begin{aligned}
({\bf P1}) ~~& \max_{\mathcal{G}}~\mathbbm{E}_{\mathcal{G}}  \left[\sum_{t=0}^{\infty}\sum_{i=1}^{N}\beta^t R(s_{i,t}, u_{i,t}) \right]  \\
&  s.t. ~~~~ \sum_{i=1}^{N}u_{i,t} = M, ~ \forall t,
\label{eq:objective}
\end{aligned}
\end{equation*}
where $\beta~(0 < \beta \le 1)$ is the discount factor. The solution to ${\bf P1}$ forms the offloading policy that determines which users are selected to offload in each time slot. 

Note that for the task offloading problem under consideration, in every slot, the user task continues to move one slot closer to their deadline, whether or not the task is offloaded in that slot. This makes the task offloading problem a restless bandit one. 
However, the formulated RMAB is a PSPACE-hard sequential decision-making problem, which is intractable in general \cite{papadimitriou1999complexity}. The complexity in deriving the optimal solution is exponential with the number of users. This undesirable condition is further exacerbated by the extremely large dimension of the system state space. Therefore, the development of a scalable and low-complexity solution enabling fast and effective user selection at each time slot is a compelling necessity.

\section{Whittle index Based Task offloading Policy}
\label{sec: WI}
Mathematically, WI \cite{whittle1988restless} provides a potential avenue to obtaining an asymptotically optimal solution to a class of RMABs with the knowledge of reward and state information. The key idea is to decouple the arms through Lagrangian relaxation, and then prove that each arm is indexable. On this basis, a complex $N$-dimensional problem can be translated into $N$ independent $1$-dimensional ones, resulting in a scalable solution with a significant reduction in the computational complexity. This motivates us to exploit the WI theory to solve the formulated RMAB when the user offloading energy consumption are available. However, the major challenge lies in how to establish the indexability (existence) and derive the WI in an easily computed form (complexity in computation). 

In this section, we first rigorously establish the indexability of the RMAB by considering a single arm reward maximization. Based on the induction method, we then prove that the formulated RMAB can admit a simple WI with closed-form expression. Finally, we elaborate on the practical implementation of the proposed policy.

\subsection{Whittle Relaxation}
A promising method, known as the Whittle relaxation, replaces the hard constraint $\sum_{i=1}^{N}u_{i,t} = M$ in $({\bf P1})$ by a soft one 
\begin{equation}
\mathbbm{E}_{\mathcal{G}} \left[\sum_{t=0}^{\infty} \beta^{t} \sum_{i=1}^{N} u_{i,t}\right] = \frac{M}{1-\beta},
\end{equation}
which only requires that the expected  discounted number of selected arms is equal to $M$. In other words, the number of selected arms at each time slot can be larger or less than $M$. In this case,  the relaxed RMAB can be shown as
\begin{equation*}
\begin{aligned}
({\bf P2})~~ &  \max_{\mathcal{G}}~\mathbbm{E}_{\mathcal{G}}  \left[\sum_{t=0}^{\infty}\sum_{i=1}^{N}\beta^t R(s_{i,t}, u_{i,t}) \right] \\
&  s.t. ~~~~ \mathbbm{E}_{\mathcal{G}} \left[\sum_{t=0}^{\infty} \beta^{t} \sum_{i=1}^{N} u_{i,t}\right] = \frac{M}{1-\beta}.
\label{eq:relaxed equation}
\end{aligned}
\end{equation*}
Leveraging the Lagrangian method, we can rewrite ${\bf P2}$ as the following unconstrained problem
\begin{equation}
\begin{aligned}
\max_{\mathcal{G}}  \mathbbm{E}_{\mathcal{G}}  \left\{\sum_{t=0}^{\infty} \left[ \sum_{i=1}^{N}\beta^t R(s_{i,t}, u_{i,t}) - \right. \right.
\hspace{2cm} \\ \left. \left. \delta \beta^t\left(\sum_{i=1}^{N} u_{i,t} - \frac{M}{1-\beta}\right)\right] \right\},
\label{eq:unconstrined}
\end{aligned}
\end{equation}
where $\delta$ is the Lagrange multiplier and will be referred to as subsidy hereafter. At this point, \eqref{eq:unconstrined} can be readily decoupled into $N$ subproblems (one for each arm) given by 

\color{black}
\begin{equation}
\begin{aligned}
& \max_{\mathcal{G}} \mathbbm{E}_{\mathcal{G}}  \left\{\sum_{t=0}^{\infty} \beta^t \left[ R(s_{i,t}, u_{i,t}) - \delta \left(u_{i,t} - \frac{M}{1-\beta}\right)\right] \right\}, \forall i. 
\\
& =  \max_{\mathcal{G}} \mathbbm{E}_{\mathcal{G}}  \left\{\sum_{t=0}^{\infty} \beta^t \left[ R(s_{i,t}, u_{i,t}) - \delta u_{i,t} \right] + \delta  \frac{M}{1-\beta} \right\}, \forall i. 
\label{eq:single unconstrained}
\end{aligned}
\end{equation}
It is clear that the $N$ separate optimization problems interact with each other \eqref{eq:single unconstrained} through a scalar Lagrange multiplier $\delta$. Taking a close look at \eqref{eq:single unconstrained} and neglecting the last constant term 
$\delta \frac{M}{1-\beta}$, our objective for each single arm $i$ is to maximize the following objective
\begin{equation}
    \max_{\mathcal{G}} \mathbbm{E}_{\mathcal{G}}  \left\{\sum_{t=0}^{\infty} \beta^t \left[ R(s_{i,t}, u_{i,t}) - \delta u_{i,t} \right] \right\}, \forall i.
\end{equation}
Then, following \cite{whittle1988restless}, we can define a modified reward of this single arm system as an equivalence to $\left[R(s_{i,t}, u_{i,t}) - \delta u_{i,t}\right]$ as follows
\begin{equation}
R^{\delta}(s_{i,t}, u_{i,t}) = R(s_{i,t}, u_{i,t}) + \delta \mathbbm{1}(u_{i,t}=0),
\label{eq:modified reward}
\end{equation}
\color{black}
where the indicator function $\mathbbm{1}(\cdot)$ gives $1$ if $u_{i,t}=0$.
We can interpret \eqref{eq:modified reward} as follows: (a) we select an arm and obtain an immediate reward $R(s_{i,t}, u_{i,t})$; (b) if the arm is not selected, we do not obtain an immediate reward (i.e., $R(s_{i,t}, u_{i,t})=0$) but receive an immediate subsidy $\delta$ (a virtual compensation from the economic view \cite{whittle1988restless}). 

Given the initial state  $s_{i,0}$, we use $V_{i,\beta}^{\delta}(s_{i,0})$ to denote the value function that represents the maximum expected total discounted reward with subsidy $\omega$. From the Bellman equation \cite{sutton1998introduction} we have 
\begin{equation}
V_{i,\beta}^{\delta}(s_{i,0}) = \max_{u_i \in \left\{ 0,1 \right\}} \left\{ R^{\delta}(s_{i,0}, u_{i}) + \beta Q_{i, \beta}^{\delta} (s_{i,0}, u_{i}) \right\}.
\end{equation}
Here, $Q_{i, \beta}^{\omega} (s_{i,t}, u_{i,t})$ is defined as
\begin{equation}
Q_{i, \beta}^{\delta} (s_{i,t}, u_{i,t}) \triangleq \sum_{s_{i,t+1}^{\prime} \in {\bf{S}}_i} p(s_{i,t+1}^{\prime} |s_{i,t}, u_{i,t})V_{i,\beta}^{\delta}(s_{i,t+1}^{\prime}),
\end{equation}
where $p(s_{i,t+1}^{\prime} |s_{i,t},u_{i,t})$ is the state transition probability from the current state $s_{i,t}$ to the next state $s_{i,t+1}^{\prime}$ given action $u_{i,t}$.
We use $\mathcal{I}(\delta)$ to represent the set of states where the optimal action $u_i^{\star}$ is not selecting the $i$-th arm, i.e.,
\begin{equation}
\mathcal{I}(\delta) \triangleq \left\{s_i:u_i^{\star}(s_i) = 0\right\}.
\end{equation}
	

\subsection{Whittle Index Based Policy}
\label{Subsection: Indexability and Whittle index}
Now we can formally introduce the concept of indexability and WI.
	
\begin{definition}[Indexability\cite{whittle1988restless}]
	The $i$-th arm is indexable if, as $\delta$ increases from $-\infty$ to $\infty$, $\mathcal{I}(\delta)$ expands monotonically from empty to the entire space. The RMAB problem is indexable if every arm is indexable.
\end{definition}
	%
	Essentially, the existence of indexability means that there is a priority order on each arm state $s_{i,t}$ in \eqref{eq:state}. Accordingly, when linking $\delta$ to $s_{i,t}$, the WI $\omega_i(s_{i,t})$, to be defined shortly, is used to quantify this order. 
	

	
\begin{theorem}
	The task offloading policy design in the MEC system formulated by the RMAB is indexable. 
\end{theorem}
Given the definition of the indexability, we now prove the indexability of the formulated RMAB. The detailed proof can be found in Appendix A.
If the indexability holds, we can assign a WI $\omega_i(s_{i,t})$ for $s_{i,t}$ to measure the criticality of each task (user), which severs as the core indicator for the user task offloading selection. The formal definition of the WI can be provided as follows.  
\begin{definition}[Whittle index\cite{whittle1988restless}]
	If an indexable arm $i$ is in state $s_{i,t}$ at the $t$-th time slot, its WI $\omega_i(s_{i,t})$ is the least value of $\delta$ for which it is optimal to make the arm passive, that is 
	\begin{equation}
	\begin{split}
	& \omega_i(s_{i,t}) \triangleq \\  & \inf_{\delta} \left\{ 
	R(s_{i,t},0) + \delta + \sum_{s_{i,t+1}^{\prime} \in {\bf{S}}_i}\beta p(s_{i,t+1}^{\prime} |s_{i},0)V_{i, \beta}^{\delta}(s_{i,t+1}^{\prime}) \right. \\ & \ge  \left. R(s_{i,t},1) + \sum_{s_{i,t+1}^{\prime} \in {\bf{S}}_i}\beta p(s_{i,t+1}^{\prime} |s_{i},1) V_{i, \beta}^{\delta}(s_{i,t+1}^{\prime}) \right\},
	\end{split}
	\end{equation}
	where $p(s_{i,t+1}^{\prime} |s_{i,t},u_{i,t})$ is the state transition probability from the current state $s_{i,t}$ to the next state $s_{i,t+1}^{\prime}$ given action $u_{i,t}$.
\end{definition}
	
After establishing the indexability of the RMAB and providing the definition of the WI, the remaining problem is how to compute the WI, which usually proves very difficult. For our RMAB, as the deadline and workload information become available once a task arrives, the arm state can be accurately captured. In the following, we will show that the unique structure of the RMAB can result in a closed-form expression for the WI.   
	
\begin{figure*}[t]
\centering
{\includegraphics[width=0.9\textwidth]{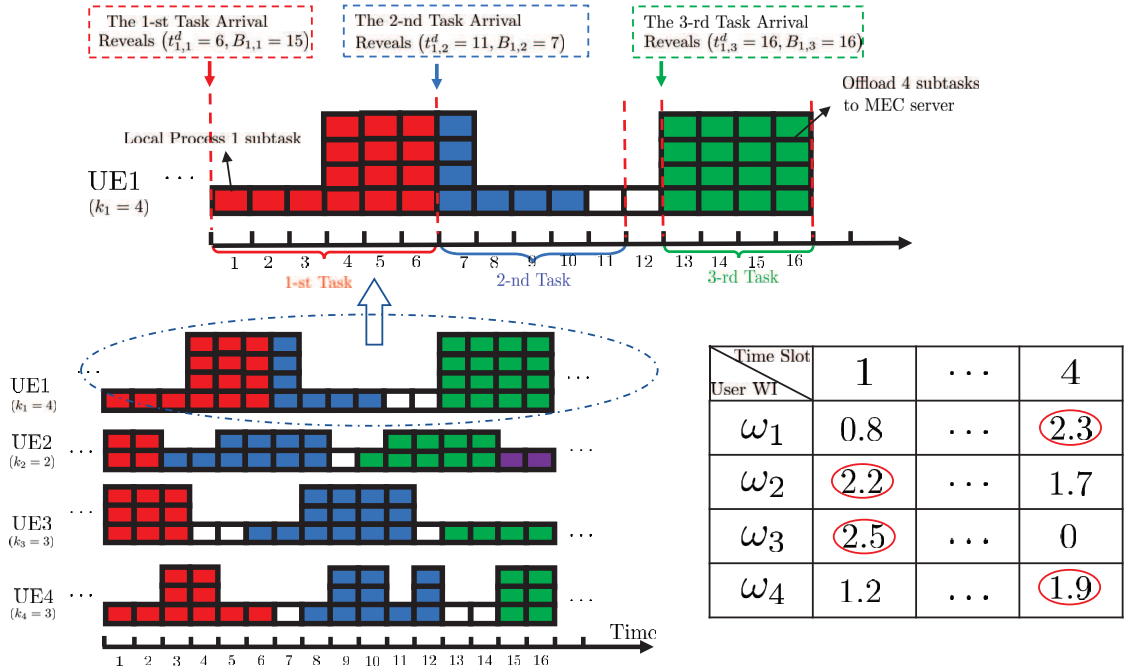}}
\caption{An example of the proposed WI policy for a large-scale asynchronous MEC system.}
\label{fig: offloading result}
\end{figure*}

\begin{theorem}
	The closed-form expression for the WI of the $i$-th user with task state $s_{i,t}=(\tau_{i,t},b_{i,t})$ is calculated as 
	\begin{equation}
	\begin{aligned}
	& \omega_i(\tau_{i,t},b_{i,t})  \\
	& =
	\begin{cases}
	0, \hspace{2.5cm}	\mbox{if } b_{i,t} = 0; 
	\\
	E_{i,j}^{\text{sav}},  \hspace{2.0cm} \mbox{if } 1 \le b_{i,t} \le (\tau_{i,t} -1) k_{i} + 1;
	\\
	E_{i,j}^{\text{sav}} + \beta^{\tau_{i,t}-1}F(b_{i,t}-k_{i}\tau_{i,t} + k_{i} - 1), 
	\\
	\hspace{2.7cm} \mbox{if } k_{i} \tau_{i,t} - k_{i} + 2 \le b_{i,t} \le k_{i} \tau_{i,t}; 
	\\
	E_{i,j}^{\text{sav}} + \beta^{\tau_{i,t}-1}F(b_{i,t}-k_{i}\tau_{i,t}+ k_{i} -1) \\
	~~- \beta^{\tau_{i,t}-1}F(b_{i,t}-k_{i}\tau_{i,t}), 
	\\
	\hspace{2.7cm} \mbox{if } b_{i,t} \ge k_{i} \tau_{i,t} + 1.
	\end{cases}
	\label{eq: Closed form Whittle}
	\end{aligned}
	\end{equation}
	\label{theorem2}
\end{theorem}
	
\begin{proof}
Please find the proof of Theorem 2 in Appendix B.
\end{proof}

\color{black}
Here, we provide some insights behind \eqref{eq: Closed form Whittle}. 
\begin{itemize}
    \item If $b_{i,t} = 0$, it means that the user has no task to offload. The WI is equal to $0$ which is also the minimal value.
    \item If $1 \le b_{i,t} \le (\tau_{i,t} - 1)k_{i} + 1$, it means that the $i$-th user's task can be finished at least one time slot ahead of the deadline. The WI is equal to the $i$-th user's energy saving $E_{i,j}^{\text{sav}}$. 
    \item If $k_{i} \tau_{i,t} - k_{i} + 2 \le b_{i,t} \le k_{i} \tau_{i,t}$, it means that the user should always be selected across all the time slots to finish its task. The WI takes into account both the energy savings and the non-completion penalty $F(b_{i,t}-k_{i}\tau_{i,t}+ k_{i} -1)$.
    \item  Finally, when it is impossible to finish the task (i.e., $b_{i,t} \ge k_i \tau_{i,t} + 1$), the WI is decreased by subtracting an extra non-completion penalty $F(b_{i,t}-k_{i}\tau_{i,t})$.
    \item Note that the selection of tasks (users) depends on the penalty parameter $\alpha$. If we focus on task completion ratio by setting a large $\alpha$, the penalty term will dominate the WI in (17). In this case, those tasks with urgent deadline (i.e.,  $k_{i} \tau_{i,t} - k_{i} + 2 \le b_{i,t} \le k_{i} \tau_{i,t}$) are given higher priority. On the other hand, when we focus on the energy consumption by setting a small $\alpha$, those tasks with higher energy savings $E_{i,j}^{\text{sav}}$ have higher priority.
\end{itemize}

\color{black}

\subsection{Implementations}
\textbf{Theorem 2} indicates that the developed task offloading  can be implemented in a very efficient manner. The whole procedure in its entirety can be summarized as follows. 
\begin{enumerate}
	\item At each $t$-th time slot, following \eqref{eq: Closed form Whittle} each user first calculates its WI $\omega_i(\tau_{i,t}, b_{i,t})$ based on the task state $\tau_{i,t}$, $b_{i,t}$ and energy saving $E_{i,j}^{\text{sav}}$. 
	\item In the message-passing phase prior to the computation offloading, each user then reports its WI $\omega_i(\tau_{i,t}, b_{i,t})$ to the BS.  As $\omega_i(\tau_{i,t}, b_{i,t})$ is a scalar, the communication overhead is quite small.  
	\item Finally, the BS selects $M$ users with the largest WI for task offloading. 
\end{enumerate}

\color{black}
An example is provided in Fig. \ref{fig: offloading result}, where we have $N=4$ users and $M=2$ MEC servers, and at each time slot only $2$ users can be selected to perform task offloading. 
Take a closer look at User 1, it has 3 tasks (indicated by red, black and green) arriving sequentially. Task $1$ with $15$ subtasks (squares) arrives at time slot $1$, revealing time slot $6$ as its deadline\footnote{The square is blank in the time slot $11$ to indicate that not task is at User 1. }.  The computation capability of User 1 and the MEC server is $1$ and $4$ subtasks per time slot, respectively\footnote{For different users, the computation capability of an MEC may different. }. 
More specifically, from time slot $1$ to time slot $3$, User 1 will process its task locally with 1 subtask per time slot as the other users (User 2 and User 3) have larger WI. Then from the time slot $4$ to time slot $6$, User 1 gets permission for task offloading, thereby $4$ subtasks are processed per time slot through offloading.

\color{black}

The developed policy features very low computational complexity and communication overhead. The key element is to calculate the WI in a distributed manner \eqref{eq: Closed form Whittle}. For the specific implementation, it takes $\mathcal{O}(N)$ time to calculate the Whittle indices of all users in Step 1). In Step 2), the sorting process has the average time complexity $\mathcal{O}(N\log (N))$. Therefore, the total time complexity is only $\mathcal{O}(N\log (N))$. The overhead of collecting Whittle indices and delivering the decision by the MEC are only $\mathcal{O}(N)$ for $N$ users.
	
\begin{remark}
Another operation is that each user reports its task state $s_{i,t}=\left(\tau_{i,t}, b_{i,t} \right)$ and the transmission cost $E_i$ to the MEC server at each time slot for the WI computation. However, this will incur the additional communication overhead. 
\end{remark}

\textcolor{black}{On the other hand, let us denote the average reward of the proposed policy, the optimal solution to $({\bf P1})$, and the solution to $({\bf P2})$ by $R_{\rm{WI}}$, $R_{\rm{Opt}}$, and $R_{\rm Relax}$, respectively. We have the following proposition to qualitatively indicate the performance of the proposed policy.}
\begin{proposition} The reward performance of the proposed policy can be shown as 
\begin{equation}
   R_{\rm{WI}}\leq R_{\rm{Opt}}\leq R_{\rm Relax}
\end{equation}
	\end{proposition}
\begin{proof}

\textcolor{black}{The first inequality naturally holds because $R_{\rm{Opt}}$ corresponds to the optimal solution to the original problem $({\bf P1})$ with hard constraints. In terms of the second inequality, note that $R_{\rm Relax}$ is the average return under the relaxed constraint in $({\bf P2})$ (does not meet the hard constraint).}
 
\end{proof}	
\textcolor{black}{
Given \textbf{Proposition 1}, it is quite difficult to quantify the gap between $R_{\rm{WI}}$ and $R_{\rm{Opt}}$.
However, we can infer their gap by numerically comparing $R_{\rm{WI}}$ with  $R_{\rm Relax}$. Due to the relationship of $R_{\rm{WI}}\leq R_{\rm{Opt}}\leq R_{\rm Relax}$ in Proposition 1, if the performance of the WI policy ($R_{\rm{WI}}$) is close to that of the relaxed policy ($R_{\rm Relax}$), we can infer that the performance gap between the proposed policy and the optimal policy $R_{\rm{Opt}}$ is very small.}

\subsection{Completion Ratio-Oriented Task Offloading Policy} 
\label{subsec: completion ratio}
In some scenarios,  we pay particular attention to the task completion ratio, i.e., the proportion of tasks that can be completed before their deadlines. 
In addition to setting a larger $\alpha$ in the penalty function, we notice that the WI \eqref{eq: Closed form Whittle} tends to give higher priority to tasks with less slack time, which is defined as $l_{i,t} \triangleq \tau_{i,t} - b_{i,t}/ k_i$ for the $i$-th user. However, the WI does not distinguish the users whose task states $s_i = (\tau_{i,t}, b_{i,t})$ satisfy $1 \le b_{i,t} \le k_i \tau_{i,t} - k_i + 1 $ (see the second case in \eqref{eq: Closed form Whittle}). This motivates us to further identify the criticality of the tasks in this scenario, thereby accommodating more tasks in a given time duration. 
	
In the reward function \eqref{eq: Reward}, we only consider the deadline breaking by imposing a penalty when $\tau_{i,t} = 1, b_{i,t} > 0$. Although considering the remaining time when $\tau_{i,t} > 1, b_{i,t} > 0$ may further reduce the risk, we may not be able to establish WI indexability and derive a very simple closed-form WI solution as \eqref{eq: Closed form Whittle}.
To address this dilemma, we proposed a priority rule referred to as shorter slack time less remaining workload (STLW).
On this basis, we propose an enhanced WI-based offloading scheduling policy by applying the STLW rule (STLW-WI). The main idea of STLW-WI is to select the users with the highest Whittle indices without violating the STLW rule. The formal definition of the STLW rule can be stated as follows.  
\begin{definition}[STLW Rule]
	\label{Definition: STLW Principle}
	Consider two users $m$ and $n$ with task states $s_{m,t}$ and $s_{n,t}$ at the $t$-th time slot. We define that the $m$-th user has priority over the $n$-th user if user $m$ has shorter slack time and less remaining workload than those of user $n$, i.e., $l_{m,t} \le l_{n,t}$ and $b_{m,t} \le b_{n,t}$, with at least one of the inequalities strictly holding.
\end{definition}
The STLW rule reorders the users based on their task states to ensure that the tasks with shorter slack time and less remaining workload should be given priority. In order to integrate the STLW rule into the WI-based offloading scheduling policy, we generate a directed acyclic graph (DAG) $\mathcal{G}=\left\{\mathcal{V},\varepsilon \right\}$, where $\mathcal{V}$ and $\varepsilon$ represent the vertex set (user set) and the edge set (users' relative priority), respectively. 
In the DAG, a directed edge from the $m$-th vertex to the $n$-th vertex indicates that the $m$-th user has the priority over the $m$-th user. \textcolor{black}{The ourdegree of a vertex $m$ is the number of directed edges leaving $m$ while the indegree of $m$ is the number of directed edges entering $m$. 
In order to preserve the priorities of users in terms of their Whittle indices whenever it is feasible, we utilize Kahn's algorithm \cite{kahn1962topological} with the largest WI vertex first criterion in the topological sorting \footnote{When no priority is set among users based on the STLW rule, users can still be ranked based on their Whittle indices}. Specifically, unlike the conventional Kahn's algorithm which selects the $0$ indegree vertex arbitrary, we select the $0$ indegree vertex with the largest WI firstly. The detailed topological sorting algorithm is shown in Algorithm~\ref{Alg: Khan's Algorithm}. 
When there is no vertex with $0$ indegree, the topological sorting is terminated, and the top $M$ users in the rank list $L_{U}$ will be chosen to offload their subtasks to the MEC server. }

\color{black}
\begin{algorithm}[!t]
	\caption{Kahn's Algorithm with largest WI vertex first.}
	\label{Alg: Khan's Algorithm}
	\begin{algorithmic}[1]
	    \color{black}
	    \STATE Initialize rank list $L_{U} = \emptyset$ that will contain the sorted user indices. 
	    \STATE Compute each vertex's indegree, i.e., the number of incoming edges for each vertex.
	    \STATE Generate a set $S$ that contains all the vertices with $0$ indegree.
          \WHILE{$S$ is not empty}
            \STATE Select the vertex $m$ in the set $S$ with largest WI and add it to the tail of the rank list $L_{U}$.
            \STATE Remove the vertex $m$ from the set DAG.
          \FOR{Each vertex $n$ with an edge from the vertex $m$ to vertex $n$}
          \STATE Decrease its indegree by $1$.
          \IF{The indegree of vertex $n$ == 0}
          \STATE Add the vertex $n$ into the set $S$.
          \ENDIF
          \ENDFOR
          \ENDWHILE
        \STATE Return rank list $L_{U}$.
	\end{algorithmic}
\end{algorithm}

	
\begin{theorem}[The STLW Performance Analysis]
	For every sequence of system state from the $t^{\prime}$-th time slot to the $(t^{\prime}+\Gamma)$-th time slot, the total discounted reward obtained by the policy with the updated STLW rule (denoted by $\mathcal{\tilde{G}}$) is not less than that achieved by the original policy (denoted by $\mathcal{G}$), i.e., we have
	\begin{equation}
	V_{\tilde{\mathcal{G}}}^{t^{\prime}+\Gamma}({\bf{S}}_{t^{\prime}}) \ge V_{\mathcal{G}}^{t^{\prime}+\Gamma}({\bf{S}}_{t^{\prime}}).
	\label{eq: STLW Performance}
	\end{equation}
	\label{theorem: STLW Performance}
\end{theorem}

\begin{proof}
Please find the proof of Theorem 3 in Appendix C.
\end{proof}

\color{black}
\begin{remark}
It is worth noting that the WI offloading policy in Section IV requires each user to report its WI at each time slot. By contrast, the STLW-WI requires each user to report its current task state $s_{i,t} = (\tau_{i,t}, b_{i,t})$.              
\end{remark}

\section{Learn to Offload Task}

\label{sec: learning}
\color{black}
In Section~\ref{Subsection: Indexability and Whittle index}, the proposed WI offloading policy requires the knowledge of user's energy saving $E_{i,j}^{\text{sav}}$ before task offloading. In some cases, however, $E_{i,j}^{\text{sav}}$ might not be available at $i$-th user prior to transmission due to lack of channel state information and offloading energy consumption $E_{i,j}^{\text{off}}$. 
In this case, the WI policy is not directly applicable (c.f. \eqref{eq: Closed form Whittle}). To address this issue, in this section, we first integrate the WI policy with the maximum likelihood estimation (MLE). To further improve the performance, we propose a novel Bayesian learning with WI policy (BL-WI) given the conjugate prior, and a refinement algorithm based on prior-swapping suitable for the non-conjugate priors (PSBL-WI).

\subsection{Maximum Likelihood Estimation with WI Policy}

Only after performing task offloading at the $t$-th time slot, the $i$-th user can obtain an estimated energy saving $e_{i,j}^{\text{sav}}$ through equipment measurement for the $j$-th task\footnote{After a successful task offloading, the user can obtain its estimated offloading energy consumption $e_{i,j}^{\text{off}}$ by the equipment measurement, and calculate its estimated energy saving by $e_{i,j}^{\text{sav}} = k_i E_i^{\text{loc}} - e_{i,j}^{\text{off}}$.}. 
Due to the energy measurement sensitivity and many other factors, $e_{i,j}^{\text{sav}}$ is a noisy version of the actual energy saving $E_{i,j}^{\text{sav}}$. Therefore, 
the observation can be written as $e_{i,j}^{\text{sav}} = E_{i,j}^{\text{sav}} + \epsilon_i$, where $\epsilon_i $ is the measurement noise. Usually the noise is the result of summing a large number of different and independent random variables. From the central limit theorem, we have $\epsilon_i \sim \mathcal{N}(0, \Sigma_i)$, where $\Sigma_i$ is the noise variance. Therefore, $e_{i,j}^{\text{sav}}$ follows a Gaussian distribution written as  $e_{i,j}^{\text{sav}} \sim \mathcal{N} \left(E_{i,j}^{\text{sav}}, \Sigma_i\right)$. 

In the following, we integrate the WI policy with the maximum likelihood estimation (MLE) technique, where we treat $E_{i,j}^{\text{sav}}$ as an unknown variable. 
Suppose that up to the $t$-th time slot for the $j$-th task offloading, the $i$-th user has performed task offloading $\gamma_{i,j}$ times, and obtained the corresponding observations $X_{i,j}(t) = \left\{e_{i,j,1}^{\text{sav}}, \cdots, e_{i, j,\gamma_{i,j}}^{\text{sav}}\right\}$.
The log likelihood is calculated by
\begin{equation}
\begin{aligned}
    \ln p \left( X_{i,j}(t) | E_{i,j}^{\text{sav}}, \Sigma_i \right) = - \frac{\gamma_{i,j}}{2} \ln(2 \pi) -  \frac{\gamma_{i,j}}{2} \ln |\Sigma_i| \\
    - \frac{1}{2} \sum_{n=1}^{\gamma_{i,j}} (e_{i,j,n}^{\text{sav}} - E_{i,j}^{\text{sav}})^2 \Sigma_i^{-1}.
\end{aligned}
\end{equation}
Taking the derivative of the log likelihood with respect to $E_{i,j}^{\text{sav}}$, we obtain
\begin{equation}
    \frac{\partial }{\partial E_{i,j}^{\text{sav}}} \ln p \left(X_{i,j}(t) | E_{i,j}^{\text{sav}}, \Sigma_i\right) = \sum_{n=1}^{\gamma_{i,j}} \Sigma_i^{-1} \left(e_{i,j,n}^{\text{sav}} - E_{i,j}^{\text{sav}}\right).
\end{equation}
By setting this derivative to zero, solution for the MLE of the energy saving is calculated as 
\begin{equation}
    \tilde{E_{i,j}^{\text{sav}}} = \frac{1}{\gamma_{i,j}} \sum_{n=1}^{\gamma_{i,j}} e_{i,j,n}^{\text{sav}}.
    \label{eq: ML for E}
\end{equation}
Clearly from \eqref{eq: ML for E}, each user can average its past observations of energy savings to obtain an estimate $\tilde{E_{i,j}^{\text{sav}}}$, and calculate its WI according to \eqref{eq: Closed form Whittle}. However, such simple update may lead to an inaccurate estimate $\tilde{E_{i,j}^{\text{sav}}}$ when the number of observations is not enough.

\subsection{Bayesian Learning with WI Policy}
To further improve the performance, we propose a novel BL-WI policy. The key is to leverage Bayesian learning to obtain the estimated energy saving $\tilde{E_{i,j}^{\text{sav}}}$ rather than just simply averaging the past observations. From the BL perspective, a prior distribution on $\tilde{E_{i,j}^{\text{sav}}}$, obtained from historical observations, can be imposed \cite{lesaffre2012bayesian}. Specifically, an observation $e_{i,j}^{\text{sav}}$ after task offloading is drawn independently from a Gaussian distribution with an unknown mean $E_{i,j}^{\text{sav}}$ and an unknown variance $\Sigma_i$, i.e., $e_{i,j}^{\text{sav}} \sim \mathcal{N} \left(E_{i,j}^{\text{sav}}, \Sigma_i\right)$. We refer to $\theta_{i,j}= \left(E_{i,j}^{\text{sav}}, \Sigma_i\right)$ as the model parameter. 

To conduct the Bayesian inference, we place a normal-inverse-gamma (NIG) conjugate prior~\cite{bishop2006pattern} on the model parameter with hyperparameters ${\lambda}_{i,j}$, $\mu_{i,j}$, $\Phi_{i,j}$ and $\nu_{i,j}$.  
In specific, the variance $\Sigma_i$ follows an inverse gamma distribution
\begin{equation}
    \Sigma_{i,j}|\left\{\Phi_{i,j},\nu_{i,j}\right\} \sim \Gamma^{-1}\left(\Phi_{i,j},\nu_{i,j}\right),
\end{equation}
and the mean $E_{i,j}^{\text{sav}}$ follows a Gaussian distribution
\begin{equation}
E_{i,j}^{\text{sav}}|\left\{\mu_{i,j},\lambda_{i,j},\Sigma_{i,j}\right\} \sim \mathcal{N}\left( {\mu}_{i,j},\frac{1}{\lambda_{i,j}}{\Sigma}_{i,j} \right).
\end{equation}
Note that the Gaussian prior is widely adopted due to a good approximation of different complex parameter distributions. 
\begin{algorithm}[!t]
	\caption{Bayesian Learning Based Whittle Index}
	\begin{algorithmic}[1]
		\STATE Initialize $\gamma_{i,j}$, model parameters $\theta_{i,j} = (\tilde{E_{i,j}^{\text{sav}}}, \tilde{\Sigma_i})$, and hyperparameters ${\lambda}_{i,j}$, $\mu_{i,j}$, $\Phi_{i,j}$, $\nu_{i,j}$,  observation sets: $X_{i,j} = \emptyset$, $\forall$ $i = 1,...,n $. 
		\FOR{$t = 0,...,T$}
        \FOR{$i = 1, \ldots, n$}
        \STATE Each user calculates its WI $\omega_i$ based on its estimated energy saving $\tilde{E_{i,j}^{\text{sav}}}$ according to \eqref{eq: Closed form Whittle}.
        \ENDFOR
        \STATE All users transmit their $\omega_i$ to the BS. 
		\STATE The BS selects the top $M$ users based on their indices ${\omega}_{i}$. Denote the selected set as $\mathcal{M}$.  
		\STATE According to the action, update each user's state according to the predefined state transition.
		\FOR{$i = 1, \ldots, n$}
		\IF{$i \in \mathcal{M}$ }
		\STATE Obtain an observation of energy saving $e_{i,j}^{\text{sav}} \sim \mathcal{N}(E_{i,j}^{\text{sav}}, \Sigma_i)$.
		\STATE  $\gamma_{i,j} = \gamma_{i,j} + 1$. 
		\STATE Append current observation into the observation set $X_{i,j}(t) \leftarrow X_{i,j}(t) \cup e_{i,j}^{\text{sav}}$.
		\STATE Update hyperparameters ${\lambda}_{i,j}$, $\mu_{i,j}$, $\Phi_{i,j}$ and $\nu_{i,j}$ according to \eqref{eq: hyperpara1}, \eqref{eq: hyperpara2}, \eqref{eq: hyperpara3}, \eqref{eq: hyperpara4}.  
		\STATE Update $\tilde{\Sigma_i}$ and $\tilde{E_{i,j}^{\text{sav}}}$ according to \eqref{eq: modelpara1} and \eqref{eq: modelpara2}.
		\ENDIF
		\ENDFOR
		\ENDFOR
\end{algorithmic}
\label{Alg: BL-WI}
\end{algorithm}
Given the observation up to the $t$-th time slot $X_{i,j}(t)$ for the $j$-th task, the $i$-th user can obtain its estimated energy saving $\tilde{E_{i,j}^{\text{sav}}}$ by efficient Bayesian inference. On this basis, we propose the BL-WI offloading policy, which is summarized in Algorithm~\ref{Alg: BL-WI}. It consists of three stages: initialization, decision making, and parameter update. 

In the initialization stage (Line 1), we initialize the model parameters, hyperparameters of NIG and counter $\gamma_{i,j}$ for each user. 
In the decision making (Lines 3-7), according to~\eqref{eq: Closed form Whittle} based on the estimated $\tilde{E_{i,j}^{\text{sav}}}$, each user calculates its WI, and then transmits it to the BS, where the $M$ users with the largest indices are selected to perform task offloading. We denote the selected user set by $\mathcal{M}$. 
In the update stage (Lines 8-17), each user first updates its task state according to the state transition defined in Section~\ref{sec: RMAB Formulation}. Then, the user in the selected set $\mathcal{M}$ increases its counter $\gamma_i$ and updates its parameters by the Bayesian inference accordingly. 
As the likelihood distribution lies in the exponential family, given the NIG prior on the unknown mean $E_{i,j}^{\text{sav}}$ and the variance ${\Sigma}_i$, we obtain the NIG posterior of $\theta_{i,j} = \left\{E_{i,j}^{\text{sav}}, {\Sigma}_i\right\} $ by the conjugacy property. Specifically, the posterior shares the same form as the prior whose hyperparameters $\lambda_{i,j}^{new}$, ${\mu}_{i,j}^{new}$, $\nu_{i,j}^{new}$, and ${\Phi}_{i,j}^{new}$ are acquired by aggregating the observations $X_{i,t}(t)$ calculated as
\begin{equation}
\lambda_{i,j}^{new} = \lambda_{i,j}^{old} + \gamma_{i,j,t},
\label{eq: hyperpara1}
\end{equation}

\begin{equation}
{\mu}_{i,j}^{new} = \frac{\lambda_{i,j}^{old}{\mu}_{i,j}^{old}+ \gamma_{i,j,t} \overline{E_{i, j,\gamma_i}^{\text{sav}}} }{\lambda_{i,j}^{old} + \gamma_{i,j,t}},
\label{eq: hyperpara2}
\end{equation}

\begin{equation}
{\Phi}_{i,j}^{new} = {\Phi}_{i,j}^{old}+\sum_{n=1}^{\gamma_{i,j,t}}\left( {e}_{i,j,n}-\overline{E_{i,j}} \right)^2 +\frac{\lambda_{i,j}^{old} \gamma_{i,j,t}}{\lambda_{i,j}^{old}+ \gamma_{i,j,t}}\frac{(\overline{E_{i,j}}-{\mu}_{i,j}^{old})^2}{2},
\label{eq: hyperpara3}
\end{equation}

\begin{equation}
\nu_{i,j}^{new} = \gamma_{i,j,t}/2 + \nu_{i,j}^{old},
\label{eq: hyperpara4}
\end{equation}
where $\overline{E_{i,j}}$ is the average of the observations $X_{i,j}(t)$. Finally, a user's estimated noise variance $\tilde{{\Sigma}_i}$ and estimated energy saving $\tilde{E_{i,j}^{\text{sav}}}$ can be sampled with updated hyperparameters as

\begin{equation}
\tilde{{\Sigma}_{i}}|\{{\Phi}_{i,j}^{new},\nu_{i,j}^{new}\} \sim \Gamma^{-1}\left( {\Phi}_{i,j}^{new},\nu_{i,j}^{new} \right),
\label{eq: modelpara1}
\end{equation}
\begin{equation}
\tilde{E_{i,j}^{\text{sav}}}|\{{\mu}_{i,j}^{new},\lambda_{i,j}^{new},{\Sigma}_{i,j}^{\prime}\} \sim \mathcal{N}\left({\mu}_{i,j}^{new},\frac{1}{\lambda_{i,j}^{new}}{\tilde{{\Sigma}_{i}}} \right).
\label{eq: modelpara2}
\end{equation}

\subsection{Refinement with BL-WI Policy}
\label{sec: learning subsection refine}
Although the NIG prior allows for a tractable and convenient Bayesian inference due to the conjugacy property, in practice, the true prior may not be conjugate (e.g., Laplace distribution). Usually, inferring the exact posterior given the non-conjugate prior is intractable, and approximate posterior inference algorithms such as Markov chain Monte Carlo (MCMC) are needed. However, inference by sampling method for the target posterior is very costly and inefficient in an online setting~\cite{zhou2018racing}, as all past observations must be involved to generate the action at each iteration.   

Motivated by~\cite{neiswanger2017post}, we adopt the prior swapping (PS) technique to make use of the pre-defined false prior (e.g., Gaussian prior), rather than running standard inference algorithms on the target prior. Here, the original Bayesian inference is divided into two simple steps: we first carry out the closed-form inference with the conjugate prior, and then utilize the PS technique to derive the posterior with the true non-conjugate prior. Hereafter, we refer to this new policy as PSBL-WI policy. 

Denote the true prior distribution over the model parameter $\theta_{i,j}$ by $\pi_t({\theta_{i,j}})$. Suppose now we have chosen a conjugate prior distribution $\pi_f(\theta_{i,j})$, which is referred to as the false prior. 
To leverage the inferred false posterior for computing the true posterior, we define a prior swapping distribution $p_s(\theta_{i,j})$  
\begin{equation}
	p_s(\theta_{i,j}) \propto \frac{ \tilde{p_f(\theta_{i,j})} \pi_t(\theta_{i,j}) }{ \pi_f(\theta_{i,j}) }, 
\end{equation} 
where $\tilde{p}_f(\theta_{i,j})$ is the interference result of the false posterior. Note that in our case, $\tilde{p}_f(\theta_{i,j}) = p_f(\theta_i|X_{i,j}(t))$ has an analytic form due to the conjugacy property, and $p_s(\theta_{i,j}) = p(\theta_i|X_{i,j}(t))$ becomes the true posterior density function. 
Then our strategy is to use $p_s(\theta_{i,j})$ in random walk Metropolis-Hastings (MH) algorithm~\cite{chib1995understanding} to approximate the true posterior distribution. Unlike the traditional MH whose computational complexity highly depends on the number of observations, the PS technique ensures that each iteration only requires to evaluate a few simple analytic expressions, and the complexity is independent of the number of observations. We denote the proposal distribution by $q (\theta_{i,j,p} | \theta_{i,j,k-1})$, where $\theta_{i,j,p}$ is the proposed sample and $\theta_{i,j,k-1}$ is the old one.
Then the MH ratio (acceptance ratio) is calculated as $\min \left(1,\rho\right)$ with
\begin{equation}
\rho = \frac{ p_s(\theta_{i,j,p}) q(\theta_{i,j,k} | \theta_{i,j,p}) }{p_s(\theta_{i,j,k}) q(\theta_{i,j,p}| \theta_{i,j,k})}.
\label{eq: MH accept ratio}
\end{equation}
Finally, after drawing $K$ samples of model parameters $\theta_{i,j}$, we can average $K$ energy saving samples $\left\{\hat{E}_{i,j,1}, \cdots, \hat{E}_{i,j,K} \right\}$ to obtain the estimated energy saving $\tilde{E_{i,j}^{\text{sav}}} = \frac{1}{K} \sum_{k=1}^{K}{\hat{E}_{i,j,k}}$. The detailed PSBL-WI policy is presented in Algorithm 3. 

\begin{algorithm}[!t]
	\caption{Prior swapping Bayesian learning Whittle index}
	\begin{algorithmic}[1]
	    \color{black}
		\STATE Initialize $\gamma_{i,j}$, model parameters $\theta_{i,j} = \left( \tilde{E_{i,j}^{\text{sav}}}, \tilde{\Sigma_i} \right)$, observation sets: $X_{i,j} = \emptyset$, $\forall$ $i = 1,...,n $, and desired number of samples $K$.
		\FOR{$t = 0,...,T$}
		\FOR{$i = 1, \cdots, n$}
		\FOR{$j = 1,\cdots, K$}
		\STATE Sample a new proposal $\theta_{i,j,p} \sim q \left(\theta_{i, j,p} | \theta_{i, j,k-1}\right)$
		\STATE Draw $\tilde{u} \sim U(0,1)$
		\IF{$\tilde{u} < \min(1, \rho)$}
		\STATE accept the proposal $\theta_{i,j,k} \leftarrow \theta_{i,j,p}$
		\ELSE
		\STATE Reject the proposal $\theta_{i,j,k} \leftarrow \theta_{i,j,k-1}$
		\ENDIF
		\ENDFOR
		\STATE Average samples $\tilde{E_{i,j}^{\text{sav}}} = \frac{1}{K} \sum_{k=1}^{K} \hat{E}_{i,j,k}$
		\STATE Each user calculates its WI $\hat{\omega}_i$ as \eqref{eq: Closed form Whittle}.
		\ENDFOR
		\STATE Same decision stage as Algorithm.~\ref{Alg: BL-WI}
		\STATE Same update stage as Algorithm.~\ref{Alg: BL-WI}.
		\ENDFOR
	\end{algorithmic}
\label{Alg: PSBL-WI}
\end{algorithm}


\color{black}
\section{Numerical Results}
\label{sec: Numerical Results}
In this section, we evaluate the performance of the proposed index-based policies by simulation. In Section~\ref{simulation: first subsection}, we first verify the proposed WI policy when the users have exact information about their energy savings from the task offloading. Then we show the impact of penalty parameter $\alpha$ on the total energy saving and completion ratio. Finally, we verify the performance of the proposed STLW-WI policy when the completion ratio becomes the main performance metric. In Section~\ref{simulation: third subsection}, we present the performance of the Bayesian learning-enabled WI policies when $E_{i,j}^{\text{sav}}$ is not available before transmission, together with its comparison to the WI policy with the knowledge of $E_{i,j}^{\text{sav}}$. 

\textcolor{black}{
The common parameters in the simulations are summarized as follows. The total rounds of task offloading is $T = 200$ with discount factor $\beta = 0.99$. The users are randomly located in the MEC system, with distance to the BS $d_{i}$ independently drawn from a uniform distribution $\mathcal{U}(0.1, 0.3)$ in kilometers. The small-scale fading channel power gains are exponentially distributed with unit mean, i.e., $\kappa_{i,j} \sim \text{Exp}(1)$~\cite{8016573}. 
We set $\sigma_0^2 = -174$ dbm/Hz, $g_0 = -40$ dB, $d_0 = 1$ m, and $\iota = 4$. Without loss of generality, each allocated sub-channel has the same bandwidth $W_i = 1$ MHz. The transmission power for each user follows a uniform distribution $P_i^{tx} \sim U(20, 25)$ dbm. For local computing, the power efficient is $\lambda = 10^{-28}$. The CPU frequency of each user $U_i$ is selected from the set $\left\{0.2, 0.4 \cdots, 1\right\}$ GHz. The required number of CPU cycles per bit is $C_i \in \left\{1, 2, 3, 4, 5\right\} \times 10^5$ cycles/bit. 
The CPU frequencies of all MEC servers are fixed $U_s = 2$ GHz. When the user is idle, the task generation probability is $Q = 0.7$. For the task specification, the duration and size of a task are bounded by $10$ time slots and $30$ subtasks, respectively. The size of a subtask is $l_i \in \left\{100, 150, 200\right\}$ bits.  The penalty function in the reward function is set as $P(b) = \alpha + 0. 1b^2$ with a different value of $\alpha$ in different subsections.  }

\begin{figure*}
	\centering
	\subfigure[Constant $M/N = 0.3$]{
		\includegraphics[width=0.45\textwidth]{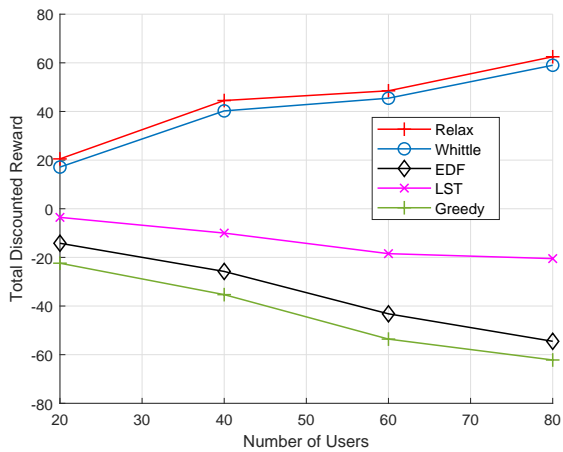}
		\label{fig: Constant M/N(a)}
	}
	\subfigure[Constant $M/N = 0.5$]{
		\includegraphics[width=0.45\textwidth]{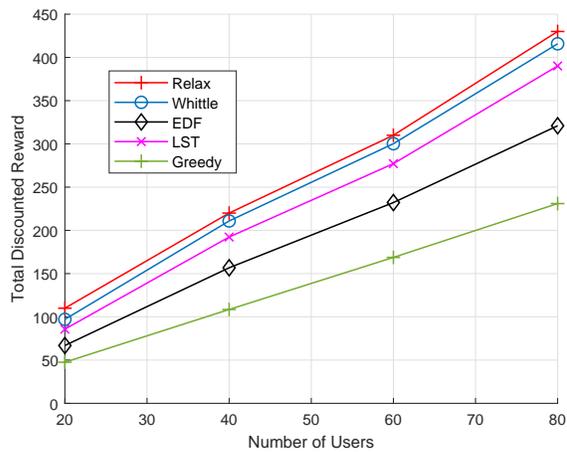}
		\label{fig: Constant M/N(b)}
	}
	\caption{Performance comparison in terms of the total discounted reward.}
	\label{fig: Constant M/N}
\end{figure*}

\subsection{The Performance of the WI Policy}
\label{simulation: first subsection}
When the perfect knowledge of $E_{i,j}^{\text{sav}}$ is available at the $i$-th user, we compare our WI policy with the following conventional policies.
\begin{itemize}
	\item \textit{Earliest Deadline First (EDF)~\cite{liu1973scheduling}:} EDF is a traditional dynamic priority policy where at each time slot the BS always selects $M$ users with minimal task remaining time $\tau_{i,t}$. Each user needs to report its current remaining time $\tau_{i,t}$ to the BS. 
	\item \textit{Least Slack Time (LST)~\cite{393496}:} LST chooses $M$ users based on their task slack time  $l_{i,t} = \tau_{i,t} - b_{i,t}/k_i$. At each time slot, the user needs to transmit its task slack time $l_{i,t}$ to the BS.
	\item \textit{Greedy Policy:} The greedy policy selects users according to their immediate reward $R_i$ as defined in \eqref{eq: Reward}, in which $M$ users with the highest rewards are selected. Each user calculates its immediate reward $R_i$ and transmits it to the BS. 
	\item \textit{Relax Solution:} In addition, we obtain the unrealistic relax solution to (${\bf P2}$) according to the method provided in~\cite{whittle1988restless}. Note that this relaxed solution is the optimal solution to {\bf{P2}}. The maximal expected average reward under relaxed constraint is 
	\begin{equation}
	    \bar{R} = \inf_{\delta} \left\{\sum_{i=1}^{N} V^{\delta}_{i,\beta} - \delta(N-M)\right\},
	\end{equation}
	where $V^{\delta}_{i,\beta}$ is the value function of the $i$-th user with subsidy $\delta$, and $\delta$ can be obtained by an exhaustive search to maximize the $\bar{R}$. Note that this solution does not satisfy the constraint in ${\bf P1}$.
\end{itemize}

To take into account both task deadline and user offloading energy consumption, we set $\alpha = 0.5$.
In comparing the WI policy with the four methods aforementioned, Fig.~\ref{fig: Constant M/N} considers two scenarios with different $M/N$. The total discounted reward is served as the performance metric. It is clearly shown that our WI policy significantly outperforms the other heuristic policies. Taking a closer look at the WI policy and the relaxed solution to $({\bf P2})$, we can infer that the performance of the WI policy is close to the optimal solution according to \textrm{Proposition 1}. 
In Fig.~\ref{fig: Constant M/N(a)}, when the number of MEC servers is not relatively enough to the number of users, the total discounted rewards obtained by heuristic policies decrease with the increment of the number of users. While our WI policy can not only achieve a positive reward but also increase with the number of users. It is because our WI policy can fully utilize the system's state information to rank the priorities among users. Compared with Fig.~\ref{fig: Constant M/N(a)} and Fig.~\ref{fig: Constant M/N(b)}, one can see that the total discounted reward increases with the ratio of available MEC servers increasing from $0.3$ to $0.5$. The reason is that more users can be selected to perform task offloading, thereby more tasks can be finished before the deadline.

\begin{figure}[t]
	\centering{
		\includegraphics[width=0.45\textwidth]{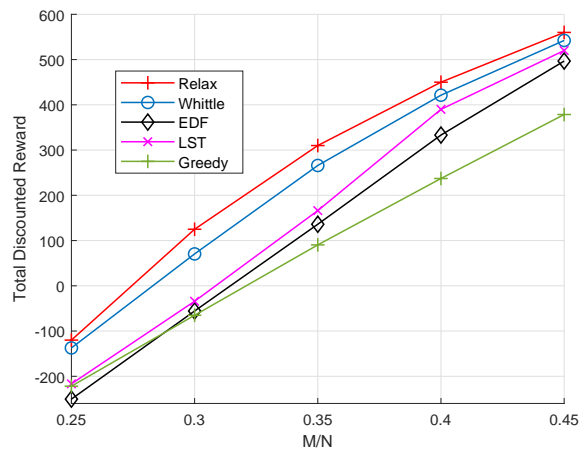}
	}
	\caption{Performance comparison in terms of the total discounted reward with constant $N$.}
	\label{fig: Constant N}
\end{figure}

In Fig.~\ref{fig: Constant N}, we fix the number of users $N=100$ and vary the number of MEC servers $M$. It can be seen that our WI policy outperforms the other policies in terms of the total discounted reward. When the number of available MEC servers is limited (e.g., $M = 25$ with $M/N = 0.25$), the performance gap between different policies is small due to very insufficient computational resources. In fact, there are a large number of tasks that cannot meet their deadlines. With the increasing number of $M$, the performance of all the policies can be improved. Given the adequate computing resources (e.g. $M = 45$ with $M/N = 0.45$), most of the tasks can be accomplished by their deadline in those policies. Therefore, all policies achieve closer performance.

\begin{figure}[t]
	\centering{
		\includegraphics[width=0.45\textwidth]{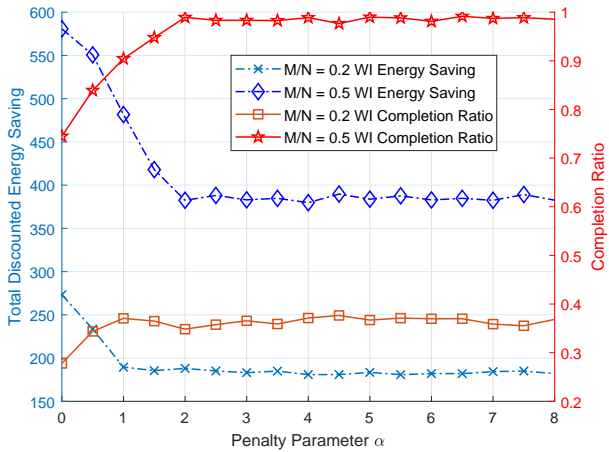}
	}
	\caption{The total discounted energy savings and completion ratio versus $\alpha$ in penalty function.}
	\label{fig: Various alpha}
\end{figure}

\begin{figure}[t]
\centering{
	\includegraphics[width=0.45\textwidth]{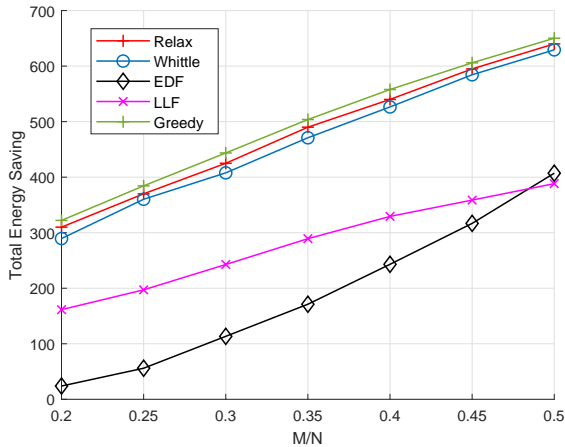}
}
\caption{Performance comparison for energy saving focus case.}
\label{fig: energy saving focus}
\end{figure}

\begin{figure*}
	\centering
	\subfigure[Completion Ratio]{
		\includegraphics[width=0.45\textwidth]{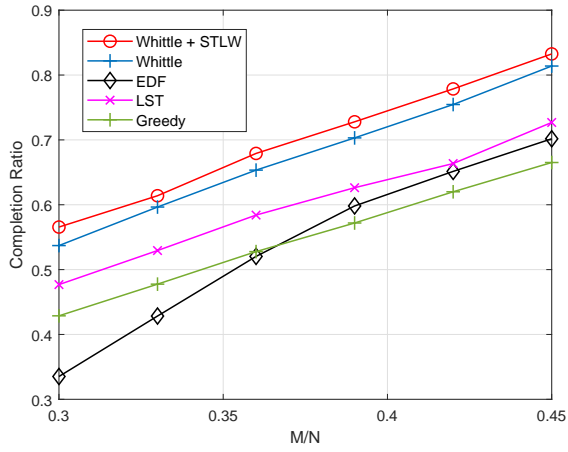}
		\label{fig: deadline focus completion ratio}
	}
	\subfigure[Reward Per Task]{
		\includegraphics[width=0.45\textwidth]{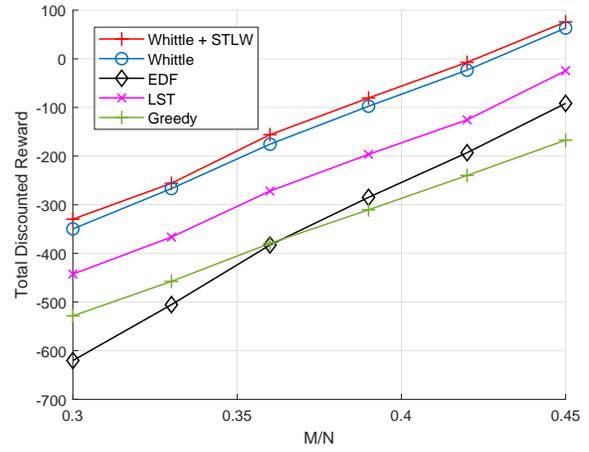}
		\label{fig: deadline focus reward}
	}
	\caption{Performance comparison for task completion focus case.}
	\label{fig: deadline focus}
\end{figure*}

 \begin{figure}[t]
	\centering
	{\includegraphics[width=0.45\textwidth]{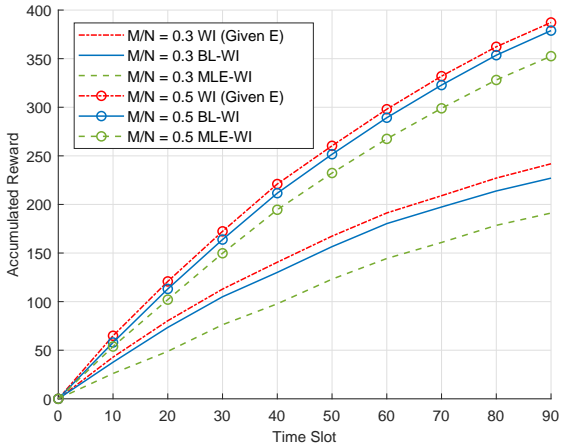}}
	\caption{The performance comparison under Gaussian prior.}
	\label{fig: Bayesian Learning Performance}
\end{figure}

 \begin{figure}[t]
	\centering
	{\includegraphics[width=0.45\textwidth]{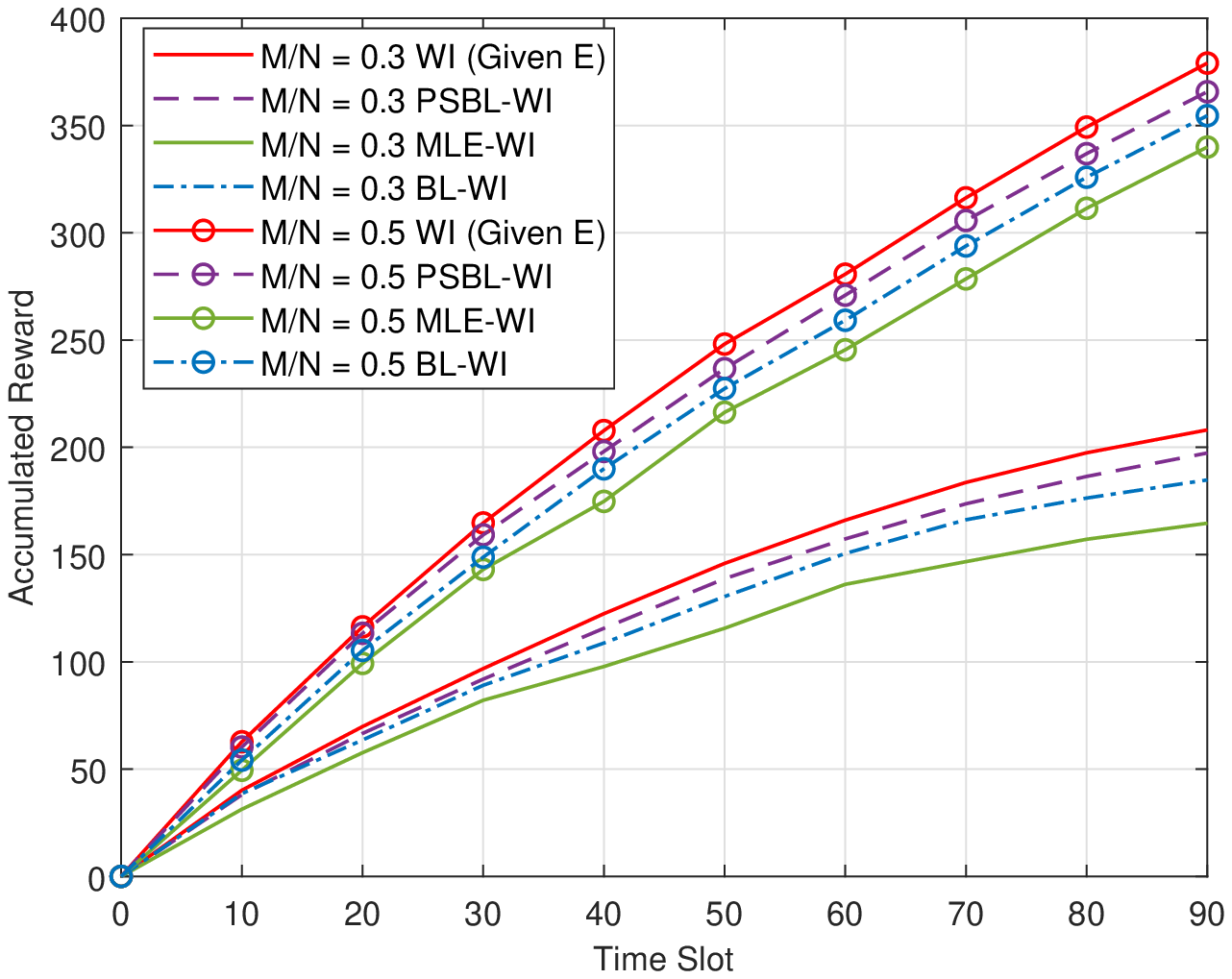}}
	\caption{The performance comparison under Laplace prior.}
	\label{fig: Bayesian Learning Performance Non-Conjugate}
\end{figure}

Essentially, the penalty parameter $\alpha$ in \eqref{eq: Reward} strikes a tradeoff between the energy saving and the task completion ratio, which is illustrated in Fig.~\ref{fig: Various alpha}. We can find that, when $\alpha$ is small, the WI policy poses an emphasis on the energy savings, resulting in a relatively low task completion ratio. With the increase of $\alpha$, the WI policy becomes task completion ratio-oriented, leading to reduced energy savings and higher task completion ratio. It is worth noting that more computational resources result in a larger $\alpha$ to make the WI policy focus on the task complete ratio. 

\textcolor{black}{In Fig.~\ref{fig: energy saving focus}, we compare the performance achieved by different policies in terms of the total energy saving with a small penalty parameter $\alpha = 0.001$. As the greedy policy only selects the task with largest energy savings to offload, it achieves maximum total energy savings. It is clearly shown that the WI-based offloading policy outperforms the EDF and LST policies, and is close to the greedy one.}
In Fig.~\ref{fig: deadline focus completion ratio}, we compare the performance achieved by different policies with a large penalty parameter $\alpha = 5$ in terms of the task completion ratio. 
Note that as the computational resource here is limited ($M < N$), not all the tasks can be finished before their deadlines.
It is clearly shown that the WI-based offloading policy outperforms the EDF, LST, and Greedy policies. Furthermore, the STLW-WI policy outperforms the original WI policy because reordering of users with the STLW rule gives urgent tasks higher priorities. When $M/N = 0.45$, the completion ratio of the proposed STLW-WI and WI  policies is 82\% and 80\%, compared to 72\%, 70\%, 66\% in LST, EDF and Greedy, respectively. The total discounted reward in Fig.~\ref{fig: deadline focus reward} also demonstrates that applying STLW rule can reduce the penalty of unfinished tasks and improve the performance. This is consistent with the theoretical analysis in Theorem~\ref{theorem: STLW Performance}.

\subsection{The Performance of the BL-enabled WI Policy}
\label{simulation: third subsection}
Next, we evaluate the performance of the proposed BL-enabled WI policy without the knowledge of user energy saving before offloading. The initialized parameters in BL-WI policy are set as follows: $\lambda_{i,j} = 1$, $\mu_{i,j} = 1$, $\Phi_{i,j} = 1$, $\nu_{i,j} = 1$, $\tilde{\Sigma_i} = 0$, $\tilde{E_{i,j}^{\text{sav}}} = 1$, and $\gamma_{i,j} = 0$.  After the $i$-th user performing task offloading at the $t$-th time slot for the $j$-th task, the observation of the energy saving is drawn from a Gaussian distribution: $e_{i,j,t} \sim \mathcal{N} (E_{i,j}^{\text{sav}}, \Sigma_i)$, with the observation noise drawn from a uniform distribution $\Sigma_i \sim U(0.5, 1)$. The desired number of samples in PSBL-WI policy is $K = 10$. For reference purpose, we also include the performance of the WI policy with the knowledge of the energy saving. 

When the energy saving $E_{i,j}^{\text{sav}}$ follows a Gaussian prior distribution, we evaluate the performance of Algorithm \ref{Alg: BL-WI} in Fig.~\ref{fig: Bayesian Learning Performance}. The prior distribution is $E_{i,j}^{\text{sav}} \sim \mathcal{N}\left(E_{i,j}^{\text{init}}, \Sigma_0\right)$ with mean $E_{i,j}^{\text{init}} = 1$ and variance $\Sigma_{i} = 0.1$.
It is clearly shown that the BL-WI policy can learn faster and more accurate than the MLE-WI policy under various $M/N$. \textcolor{black}{Since the channel gain changes every 20 time slots, the performance of MLE-WI policy is limited by the number of observations collected in 20 time slots.}  Additionally, comparing with WI policy, the reward gaps of both BL-WI policy and MLE-WI policy decrease when the number of available MEC servers increases, i.e., $M/N$ increases from $0.3$ to $0.5$,. In particular, the BL-WI policy achieves a much more significant performance improvement than the MLE-WI policy counterpart. It is because more users have opportunities to perform task offloading and obtain more observation samples, accelerating the Bayesian learning process and  decreasing the sample bias in MLE. 

Next, we evaluate the performance of Algorithm \ref{Alg: PSBL-WI} when the energy saving $E_{i,j}^{\text{sav}}$ has a non-conjugate prior distribution. Specifically, we place a Laplace distribution as the prior distribution  $E_{i,j}^{\text{sav}} \sim Laplace\left({E}_{i,j}^{\text{init}}, b_0 \right)$ with location parameter $E_{i,j}^{\text{init}} = 1$ and scale parameter $b_0 = 0.2$. For the BL-WI policy, we still use a conjugate NIG prior to performing exact Bayesian inference. 
Fig.~\ref{fig: Bayesian Learning Performance Non-Conjugate} shows that the PSBL-WI policy outperforms the other ones in the non-conjugate case. Although the Gaussian prior brings the convenience in Bayesian inference, comparing the performance of BL-WI policy in Fig.~\ref{fig: Bayesian Learning Performance} and Fig.~\ref{fig: Bayesian Learning Performance Non-Conjugate}, the reward gap between BL-WI policy and WI policy increases due to the false prior assumption. 
Similar to the conjugate prior case, when the number of MEC servers increases, the performance of BL-enabled WI policies improve with more samples obtained during the offloading.

\subsection{Discussion}
\color{black}
In our problem, we assume a pre-allocated bandwidth scheme. However, it is possible to include the bandwidth allocation into the formulated problem. To do so, the action at each time slot will be modified as $a_{i,t} = (u_{i,t}, d_{i,t})$, where  $u_{i,t} \in \left\{0, 1\right\}$ is the offloading decision and $d_{i,t} \in \left[d_{\min}, d_{\max}\right]$ the bandwidth allocation. Accordingly, two constraints imposed on the question $\sum_{i}^{N} u_{i,t} = M $ (only $M$ users can be selected to perform task offloading), and $\sum_{i} u_{i,t} d_{i,t} = W $ (the sum of bandwidth allocation is $W$). 
This new RMAB problem with an extra constraint (bandwidth limitation) makes the establishment of the indexability difficult, and we consider this problem as our future work. 

\textcolor{black}{It is worth noting that the task offloading in a large-scale asynchronous MEC system may suffer from Byzantine failure, where 
the status of server or users appears to be in failure to some users while functional to other users. Therefore, the system needs to first reach a consensus on whether the user or server has failed, then it can shut down the failure part accordingly. To handle this problem, we may resort to the asynchronous Byzantine fault tolerant protocol proposed in \cite{miller2016honey}. 
Under this framework, users receive the offloading history from other users and
store them in their buffer. At the beginning of each epoch, each user selects and provides a subset of the history in its buffer to a randomized agreement protocol which is used to determine whether the target user is in failure.
We will investigate the specific implementation of this method as our future topic.}

\color{black}
Note that due to the limited computational resources, the task completion ratio cannot be further improved by our proposed method. In practical applications (e.g., the task offloading for non-critical wireless sensors such as smart meters reading \cite{7281870, 6574667}), we need other supplementary methods to tolerate high violation probability further. For example, if the task misses its current deadline, it can be stored in the buffer and assigned a new deadline for later task offloading. 
\color{black}


\color{black}

\section{Conclusions}
\label{section 7}
We proposed a novel WI-based task offloading policy for a large-scale asynchronous MEC system, which features scalable calculation and simple implementation. We formulated the offloading policy design as an RMAB with the objective to maximize the total discounted reward over the time horizon. Based on the WI theory, we rigorously established the indexability and derived the WI in a closed-form expression.
To achieve a higher task completion ratio, the STLW-WI policy is proposed in the task completion ratio-oriented case.  
For the case of unknown user offloading energy consumption prior to offloading, we proposed the BL-WI policy and PSBL-WI policy for the conjugate and non-conjugate prior cases, respectively.   
Simulation results verified that the proposed policies significantly outperform the existing policies.
The proposed method provides a potential avenue to the highly efficient task offloading with the upcoming large-scale MEC deployment in the IoT.

\appendices
\section{Proof Of Theorem 1}
\begin{proof}
Without loss of generality, we drop the subscript $i$, $j$, $\delta$ and $t$. Denote the difference between two value functions by $h(\tau,b) = V(\tau, b+k-1) - V(\tau, b)$, and the difference of two actions (offloading and un-offloading) by $g(\tau,b)$. The indexability of the offloading problem depends on the property that $h(\tau,b)$ is piecewise linear in $\delta$ and $\frac{ \partial h(\tau,b)}{ \partial \delta} \ge -1 $, because this property guarantees that $\frac{ \partial g(\tau,b)}{ \partial \delta} = \left[1 - \frac{ \partial h(\tau,b)}{ \partial \delta} \right] \ge 0 $. The induction method is applied to prove this property. Specifically, we first show that the WI $\omega(\tau,b) $ exists for $\tau = 0, 1$, then assuming that the WI exists and $\frac{ \partial h(\tau,b)}{ \partial \delta} \ge -1 $ for $\tau = t - 1$, we show that the WI also exists and $\frac{ \partial h(\tau,b)}{ \partial \delta} \ge -1 $ holds for $\tau = t$.


\begin{enumerate}
	\item $\tau=0$: There is no task waiting in the user. The Bellman equation is stated as 
	\begin{equation}
	V(0,0) = \max\left\{\delta+\beta V_e, \beta V_e \right\},
	\end{equation}
	where $V_e$ is the expected reward of future tasks generation. Therefore, if and only if $\delta>0$, the first term is larger and the un-offloading action is optimal. Thus $\omega(0,0) = 0$.
	\item $\tau = 1$: there are four cases.
	\begin{enumerate}
		\item If $b = 0$, the Bellman equation is stated as
		\begin{equation}
		V(1,0) = \max\left\{\delta+\beta V_e, \beta V_e \right\}.
		\end{equation}
		It is same as the case $\tau=0, b=0$, therefore, $\omega(1,0) = 0$.
		\item If $ b = 1$, the Bellman equation is stated as
		\begin{equation}
		V(1,b) = \max\left\{\delta + \beta V_e,  E^{\text{sav}} +\beta V_e \right\}. 
		\end{equation}
		If and only if $\delta \ge E^{\text{sav}}$, the un-offloading action is optimal. Thus $\omega(1,b) = E^{\text{sav}}$ when $b = 1$.
		\item If $ 1 < b \le k$, the Bellman equation is stated as
		\begin{equation}
		\begin{aligned}
		V(1,b) = \max\left\{\delta +\beta V_e - F(b-1), E^{\text{sav}} + \beta V_e \right\}. 
		\end{aligned}
		\end{equation}
		If and only if $\delta \ge E^{\text{sav}} + F(b-1)$, the un-offloading action is optimal. Thus $\omega(1,b) = E^{\text{sav}} + F(b-1)$ when $1 < b \le k$. 
		\item If $ b > k $, the Bellman equation is stated as
		\begin{equation}
		\begin{aligned}
		V(1,b) = \max\left\{\delta - F(b-1) + \beta V_e, \right.
		\\
		\left. 
		E^{\text{sav}} - F(b-k) + \beta V_e \right\}. 
		\end{aligned}
		\end{equation}
		If and only if $\delta \ge E^{\text{sav}} + F(b-1) -F(b-k)$, the un-offloading action is optimal. Thus 
		\begin{equation}
		\omega(1,b) = E^{\text{sav}} + F(b-1) -F(b-k).
		\end{equation}
	\end{enumerate}
	Thus the WI for $\tau=1$ exists, and the closed form is given by
	\begin{equation}
	\omega(1,b) = 
	\begin{cases}
	0, & \mbox{if } b = 0; \\
	E^{\text{sav}} & \mbox{if } b = 1; \\
	E^{\text{sav}} + F(b-1),  & \mbox{if } 1 < b \le k; \\
	E^{\text{sav}} + F(b-1) - F(b-k) & \mbox{if } b > k; \\
	\end{cases}
	\label{WI when tau = 1}
	\end{equation}	
\end{enumerate}

Now we are ready to show the $\frac{ \partial h^{\delta}(\tau,b)}{ \partial \delta} \ge -1 $ holds when $\tau = 1$. 
	\begin{enumerate}
		\item If $b = 0$, $h(1,0)=V(1,k-1)-V(1,0)$, we have
		\begin{equation}
		h(1,0) = 
		\begin{cases}
		E^{\text{sav}}, &  \mbox{if } \omega < 0; \\
		E^{\text{sav}}-\delta, & \mbox{if } 0 \le \delta < \omega(1,k-1); 
		\\
		-F(k-2), & \mbox{if } \delta \ge \omega(1,k-1).
		\end{cases}
		\end{equation}
		
		\item If $b = 1$, $h(1,1)=V(1,k)-V(1,1)$, we have
		\begin{equation}
		h(1,1) = 
		\begin{cases}
		0, & \mbox{if } \delta < \omega(1, 1); \\
		E^{\text{sav}} - \delta, & \mbox{if } \omega(1, 1) \le \delta < \omega(1, k); \\
		-F(k-1), & \mbox{if } \delta \ge \omega(1,k).
		\end{cases}
		\end{equation}
		
		\item If $2 \le b \le k$, $h(1,b)=V(1,b + k - 1)-V(1,b)$, we have
		\begin{equation}
		h(1,b) = 
		\begin{cases}
		-F(b-1), \\  
		\hspace{0.3cm} \mbox{if } \delta < \omega(1, b + k -1); \\
		E^{\text{sav}} - \delta,  \\  
		\hspace{0.3cm} \mbox{if } \omega(1,b) \le \delta < \omega(1, b + k -1); 
		\\
		-F(b+k-2)+F(b-1), \\ \hspace{0.3cm} \mbox{if } \delta \ge \omega(1,b).
		\end{cases}
		\end{equation}
		
		\item If $b > k$, $h(1,b)=V(1,b + k - 1)-V(1,b)$, we have
		\begin{equation}
		h(1,b) = 
		\begin{cases}
		-F(b-1)+F(b-k), 
		\\
		\hspace{0.3cm} \mbox{if } \delta < \omega(1,b); \\
		E^{\text{sav}} - \delta, \\ \hspace{0.3cm} \mbox{if } \omega(1,b) \le \delta < \omega(1, b + k -1); \\
		-F(k+b-2)+F(b-1), \\ \hspace{0.3cm} \mbox{if } \delta \ge \omega(1, b + k -1).
		\end{cases}
		\end{equation}
		
	\end{enumerate}		
	Therefore, the derivation of $h(1,b)$ on $\delta$ always guarantees that $\frac{\partial h(1,b)}{\partial\delta} \ge -1$, which implies the indexability holds when $\tau=1$.
	Then, we show the property of $h(\tau,b)$ holds when $\tau=t$ under the assumption that $\frac{\partial h(\tau,b)}{\partial \delta} \ge -1$ when $\tau = t-1$.

    \begin{enumerate}
		\item  If $b=0$, 
		\begin{equation}
		h(\tau,0) =
		\begin{cases}
		E^{\text{sav}}, \hspace{0.3cm} \mbox{if } \delta < 0; \\
		E^{\text{sav}}-\delta, \\ \hspace{0.3cm} \mbox{if } 0 \le \delta < \omega(\tau,k-1); \\
		\beta \left[V(\tau-1,k-2) - V(\tau-1,0) \right], 
	    \\ \hspace{0.3cm} \mbox{if } \delta \ge \omega(\tau,k-1).
		\end{cases}
		\end{equation}
		
		For the first two cases, it is clearly shown that the gradient of $h(\tau,b)$ with respect to $\delta$ is larger or equal than $-1$. For the third case, we can further expand it by comparing the value of $\delta$ and $\omega(\tau-t^{\prime}, k - t^{\prime}-1)$. Whenever there is a time step $t^{\prime}$ such that $\delta \le \omega(\tau-t^{\prime}, k - t^{\prime}-1)$, $\exists 2 \le t^{\prime} \le \tau$, 				$h(\tau,0) = \beta^{t^{\prime}} \left(E^{\text{sav}} - \omega \right)$ whose gradient is $-\beta^{t^{\prime}} \ge -1$. On the other hand, if there is no such time step, $h(\tau,0)$ will be calculated by its penalty term whose gradient in terms of $\omega$ is zero.
		
		\item If $ 1 \le b \le k$, $h(\tau,b) = V(\tau,b + k -1)-V(\tau,b)$, we have
		\begin{equation}
		h(\tau,b) =
		\begin{cases}
		\beta \left[V(\tau-1, b-1) - V(\tau-1, 0)\right], 
		\\
		\hspace{0.3cm} \mbox{if } \delta <\omega(\tau,b); \\
		E^{\text{sav}}-\delta, \\ \hspace{0.3cm} \mbox{if } \omega(\tau,b) \le \delta < \omega(\tau,b + k -1); \\
		\beta h\left(\tau-1, b-1\right) \\ \hspace{0.3cm} \mbox{if } \delta \ge \omega(\tau,b+ k -1).
		\end{cases}
		\end{equation}
		
		For the first case, a similar analysis as the third case in $b=0$ can be carried out here and we have $\frac{\partial h(\tau,b)}{\partial \delta}\ge -1$ as well. Since we have $\frac{\partial h(\tau-1,b-1)}{\partial \delta}\ge -1$ for all $b$ by assumption, we have $\frac{\partial h(\tau,b)}{\partial \delta}\ge -1$ here as well.

		\item If $ b > k$, $h(\tau,b) = V(\tau,b + k -1)-V(\tau,b)$, we have
		
		\begin{enumerate}
			\item If $\omega(\tau,b + k - 1) > \omega(\tau,b)$,  we have
			\begin{equation}
			h(\tau,b) =
			\begin{cases}
			\beta h(\tau-1, b-k), \\ \hspace{0.3cm} \mbox{if } \delta<\omega(\tau,b); \\
			E^{\text{sav}}-\delta, \\ \hspace{0.3cm}  \mbox{if } \omega(\tau,b) \le \delta < \omega(\tau,b + k -1); \\
			\beta h(\tau-1, b-1), \\ \hspace{0.3cm} \mbox{if } \delta \ge \omega(\tau,b+ k -1).
			\end{cases}
			\end{equation}
			
			\item If $\omega(\tau,b + k - 1) \le \omega(\tau,b)$, we have
			\begin{equation}
			h(\tau,b) =
			\begin{cases}
			\beta h(\tau-1, b-k), \\ \hspace{0.3cm} \mbox{if } \delta < \omega(\tau,b + k - 1); \\
			\delta - E^{\text{sav}} + \beta \left[h(\tau-1, b - 1) \right. \\ \left.  + h(\tau-1, b - k)\right], \hspace{0.3cm}
			\\ \hspace{0.3cm} \mbox{if } \omega (\tau,b+1) \le \delta < \omega(\tau,b); \\
			\beta h(\tau-1, b-1), \\ \hspace{0.3cm} \mbox{if } \delta \ge \omega(\tau,b).
			\end{cases}
			\end{equation}
		\end{enumerate}
	\end{enumerate}
	Since $\frac{\partial h(\tau-1,b-1)}{\partial \omega}\ge -1$ for all $b$ by assumption, we have $\frac{\partial h(\tau,b)}{\partial \omega}\ge -1$ in all cases. 
	Thus, the indexability of the RMAB can be established.

\end{proof}

\section{Proof Of Theorem 2}
\begin{proof}
In this section, we derived the closed form of the WI by induction. Since the case for $\tau = 0$ and $\tau=1$ have been proved during the indexability prove, we start from $\tau = 2$ from here. 

\begin{enumerate}
	\item $\tau = 2$: there are four cases.
	\begin{enumerate}
		\item If $b = 0$, the Bellman equation is stated as
		\begin{equation}
		V(2,0) = \max\left\{\delta+\beta V(1,0), \beta V(1,0) \right\}.
		\end{equation}
		It is same as the case $\tau=0, b=0$, therefore, $\omega(2,0) = 0$.
		
		\item If $b = 1$, the Bellman equation is stated as
		\begin{equation}
		V(2,b) = \max\left\{\delta + \beta V(1,0), E^{\text{sav}}+\beta V(1,0) \right\}. 
		\end{equation}
		If and only if $\delta \ge E^{\text{sav}}$, the un-offloading action is optimal. Thus $\omega(2,1) = E^{\text{sav}}$. 
		
		\item If $ 1 < b \le k$, the Bellman equation is stated as
		\begin{equation}
		V(2,b) = \max\left\{\delta + \beta V(1,b-1), E^{\text{sav}} +\beta V(1,0) \right\}. 
		\end{equation}
		The difference between actions is 
		\begin{equation}
		\begin{aligned}
		g(2, b) & = \delta - E^{\text{sav}} + \beta \left[V(1,b-1) - V(1,0)\right]
		\\ 
		& = 
		\begin{cases}
		\delta - E^{\text{sav}} + \beta E^{\text{sav}}  \\ \hspace{0.3cm} \mbox{if } \delta<0; \\
		\delta - E^{\text{sav}}  + \beta \left(E^{\text{sav}} - \delta \right), \\ \hspace{0.3cm} \mbox{if } 0 \le \delta \le \omega(1, b-1) ; \\
		\delta - E^{\text{sav}}  - \beta F(b-2), \\ \hspace{0.3cm} \mbox{if } \delta > \omega(1, b-1); \\
		\end{cases}	
		\end{aligned}
		\end{equation}
		The difference equals $0$ when $\delta = E^{\text{sav}} $. Thus $\omega(2,b) = E^{\text{sav}}$ when $ 1 < b \le k$.
		
		\item If $ b = k + 1$, the Bellman equation is stated as
		\begin{equation}
		V(2,k + 1) = \max\left\{\delta +\beta V(1,k), E^{\text{sav}}  +\beta V(1,1) \right\}. 
		\end{equation}
		The difference between actions is 
		\begin{equation}
		\begin{aligned}
		g(2, k + 1) & = \delta - E^{\text{sav}}  + \beta \left[V(1,k) - V(1,1)\right]
		\\ 
		& = 
		\begin{cases}
		\delta - E^{\text{sav}}, \\ \hspace{0.3cm} \mbox{if } \omega< \omega(1,1); \\
		\delta - E^{\text{sav}}  + \beta \left(E^{\text{sav}} - \delta \right), \\ \hspace{0.3cm} \mbox{if } \omega(1,1) \le \delta \le \omega(1, k); \\
		\delta - E^{\text{sav}}  - \beta F(k -1), \\ \hspace{0.3cm} \mbox{if }\delta > \omega(1, k); \\
		\end{cases}
		\end{aligned}	
		\end{equation}
		The difference equals $0$ when $\delta = E^{\text{sav}}$. Thus $\omega(1,b) = E^{\text{sav}} $ when $ b = k+1$.
		
		\item If $k + 1 < b \le 2k $, the Bellman equation is stated as
		\begin{equation}
		\begin{aligned}
		V(1,b) = \max\left\{\delta+\beta V(1,b-1), \right.
		\\
		\left. E^{\text{sav}} + \beta V(1,b-k) \right\}. 
		\end{aligned}
		\end{equation}
		
		The difference between actions is 
		\begin{equation}
		\begin{aligned}
		g(2, b) & = \delta - E^{\text{sav}} + \beta \left[V(1,b-1) - V(1,b-k)\right]
		\\ 
		& = 
		\begin{cases}
		\delta - E^{\text{sav}} + \beta \left[- F(b-k-1)\right] \\ \hspace{0.3cm} \mbox{if } \delta <\omega(1, b-k); \\
		\delta - E^{\text{sav}} + \beta \left[E^{\text{sav}} - \delta \right], \\ \hspace{0.3cm} \mbox{if } \omega(1, b - k) \le \delta \le \omega(1, b - 1) ; \\
		\delta - E^{\text{sav}} + \beta \left[-F(b-2) + F(b-k-1)\right], \\ \hspace{0.3cm} \mbox{if } \delta > \omega(1, b-1); \\
		\end{cases}
		\end{aligned}	
		\end{equation}
		The difference equals $0$ when $\delta = E^{\text{sav}} + \beta \left[ F(b-k-1)\right]$. Thus $\omega(2, b) = E^{\text{sav}} + \beta \left[ F(b-k-1)\right] $ when $ k + 1 < b \le 2k $.
		
		\item If $b > 2k $, the Bellman equation is stated as
		\begin{equation}
		\begin{aligned}
		V(1,b) = \max\left\{\delta + \beta V(1,b-1),  E^{\text{sav}} +\beta V(1,b-k) \right\}. 
		\end{aligned}
		\end{equation}
		
		
		The difference between actions is 
		\begin{equation}
		\begin{aligned}
		g(2, b) & = \delta - E^{\text{sav}}  + \beta \left[V(1,b-1) - V(1,b-k)\right]
		\\ 
		& = 
		\begin{cases}
		\delta -  E^{\text{sav}}  + \beta \left[-F(b-k-1) + F(b-2k)\right], \\ \hspace{0.3cm} \mbox{if } \delta <\omega(1, b-k); \\
		\omega -  E^{\text{sav}}+ \beta \left[E^{\text{sav}}- \delta \right], \\ \hspace{0.3cm} \mbox{if } \omega(1, b-k) \le \delta < \omega(1, b - 1) ; \\
		\omega -  E^{\text{sav}} + \beta \left[-F(b-2) + F(b-k-1)\right], \\ \hspace{0.3cm} \mbox{if } \delta \ge \omega(1, b - 1); \\
		\end{cases}
		\end{aligned}	
		\end{equation}
		The difference equals $0$ when $\delta = E^{\text{sav}} + \beta\left[F(b-k-1) - F(b-2k)\right]$. Thus $\omega(1,b) = E^{\text{sav}} + \beta\left[F(b-k-1) - F(b-2k)\right]$ when $  b > 2k $.
	\end{enumerate}

	
	Thus the WI for $\tau=2$ exists, and the closed form is calculated as
	\begin{equation}
	\omega(2,b) = 
	\begin{cases}
	0, \hspace{1.5cm} \mbox{if } b = 0; \\
	E^{\text{sav}}, \hspace{1.1cm} \mbox{if } 1 \le b \le k + 1; \\
	E^{\text{sav}} + \beta \left[ F(b-k-1)\right],  \\ \hspace{0.3cm} \mbox{if } k + 1 < b \le 2 k; \\
	E^{\text{sav}} + \beta\left[F(b-k-1) - F(b-2k)\right] \\ \hspace{0.3cm} \mbox{if } b > 2 k; \\
	\end{cases}
	\end{equation}
	
\end{enumerate}

Next we show the closed-form of the WI for the case of $\tau \ge 3$, assuming (1) holds for $\tau-1$. 

\begin{enumerate}
\item If $b = 0$, the Bellman equation is stated as
\begin{equation}
V(\tau,0) = \max\left\{\delta + \beta V(\tau-1,0), \beta V(\tau-1,0) \right\}.
\end{equation}
Therefore, $\omega(\tau,0) = 0$.
\item If $b= 1$, the Bellman equation is stated as 
\begin{equation}
\begin{aligned}
V(\tau,1) = \max\left\{\delta + \beta V\left(\tau-1,0\right), E^{\text{sav}}+\beta V\left(\tau-1, 0 \right) \right\}. 
\end{aligned}
\end{equation}
Thus $\omega(\tau,1) =  E^{\text{sav}}$.

\item If $ 2 \le b \le k$, the Bellman equation is stated as
\begin{equation}
\begin{aligned}
V(\tau,b) = \max\left\{\omega + \beta V\left(\tau-1,b-1\right), 
\right. \\
\left.
E^{\text{sav}}+\beta V\left(\tau - 1, 0 \right) \right\}. 
\end{aligned}
\end{equation}
The difference between actions is:
\begin{equation}
\begin{aligned}
g(\tau,b) & = \delta - E^{\text{sav}} + \beta \left[V(\tau-1, b-1) - V(\tau-1, 0)\right]
\\
& = 
\begin{cases}
\delta - E^{\text{sav}} + \beta^2 E^{\text{sav}}, \\ \hspace{0.3cm} \mbox{if } \delta < 0; \\
\delta - E^{\text{sav}} + \beta \left[E^{\text{sav}} - \delta \right],  \\ \hspace{0.3cm} \mbox{if } 0 \le \delta \le \omega(\tau-1, b - 1); \\
\omega - E^{\text{sav}} + \beta^2 \left[V(\tau-2, b-2) - V(\tau-2, 0)\right] \\ \hspace{0.3cm} \mbox{if }  \delta > \omega(\tau-1, b - 1); \\
\end{cases}
\end{aligned}
\end{equation}
The difference equals $0$ when $\delta = E^{\text{sav}}$. Thus $\omega(\tau,b) = E^{\text{sav}}$ when $ 2 \le b \le k$.

\item If $ k < b \le k (\tau-2) + 2$, the Bellman equation is stated as
\begin{equation}
\begin{aligned}
V(\tau, b) = \max\left\{\delta+\beta V\left(\tau-1,b-1\right), E^{\text{sav}} + \beta V\left(\tau-1,b-k\right) \right\}. 
\end{aligned}
\end{equation}


The difference between actions is
\begin{equation}
\begin{aligned}
g(\tau,b) & = \delta - E^{\text{sav}} + \beta \left[V\left(\tau-1,b-1\right)- V\left(\tau-1,b-k\right)\right]
\\ 
& = 
\begin{cases}
\delta - E^{\text{sav}} + \beta^2 \left[V\left(\tau-2,b-1-k\right)- \right. \\ \left. V\left(\tau-1,(b-2 k)^{+}\right)\right], \hspace{0.5cm} \mbox{if } \delta < E^{\text{sav}}; \\
\delta - E^{\text{sav}} + \beta^2 \left[V\left(\tau-2,b-2\right)- \right. \\ \left. V\left(\tau-2,b- k -1\right)\right],  \hspace{0.5cm} \mbox{if } \delta \ge E^{\text{sav}}; \\
\end{cases}	
\end{aligned}
\end{equation}
Since we have $\omega(\tau-1, b - k -1) = E^{\text{sav}}$ when $1 \le b - k -1 \le k(\tau-2) + 1$ 
So $\omega(\tau,b) = E^{\text{sav}}$ when $b= k+1$. 

\item If $ k (\tau-2) + 2 < b \le k (\tau-1) + 1$, the Bellman equation is stated as
\begin{equation}
\begin{aligned}
V(\tau, b) = \max\left\{\delta+\beta V\left(\tau-1,b-1\right), \right. \\ \left. E^{\text{sav}}+\beta V\left(\tau-1,b-k\right) \right\}. 
\end{aligned}
\end{equation}

Therefore, the difference between actions is
\begin{equation}
\begin{aligned}
g(\tau,b) & = \delta - E^{\text{sav}} + \beta \left[V\left(\tau-1,b-1\right)- V\left(\tau-1,b-k\right)\right]
\\ 
& = 
\begin{cases}
\delta - E^{\text{sav}} + \beta^2 \left[V\left(\tau-2,b-1-k\right)- \right. \\ \left.  V\left(\tau-1,(b-2 k)\right)\right], \hspace{0.3cm} \mbox{if } \delta < \omega(\tau-1, b - k); 
\\
\delta - E^{\text{sav}} + \beta \left[\delta - E^{\text{sav}}\right] \\ \hspace{0.3cm} \mbox{if } \omega(\tau-1, b - k) \le \delta < \omega(\tau-1, b-1);
\\
\delta - E^{\text{sav}} + \beta^2 \left[V\left(\tau-2,b-2\right)- \right. \\ \left.  V\left(\tau-2,b- k -1\right)\right],  \hspace{0.3cm} \mbox{if } \delta \ge \omega(\tau-1, b-1); \\
\end{cases}	
\end{aligned}
\end{equation}
It equals $0$ when $\delta = E^{\text{sav}}$. Thus $\omega (\tau,b) = E^{\text{sav}}$ when $ k (\tau-2) + 2 < b \le k (\tau-1) + 1$.

\item If $ k(\tau-1)+2 \le b \le k \tau$, the Bellman equation is stated as
\begin{equation}
\begin{aligned}
V(\tau, b) = \max\left\{\delta +\beta V\left(\tau-1,b-1\right), \right. \\ \left. E^{\text{sav}}+\beta V\left(\tau-1,b-k\right) \right\}. 
\end{aligned}
\end{equation}

Therefore, the difference between actions is
\begin{equation}
\begin{aligned}
g(\tau,b) & = \delta - E^{\text{sav}} + \beta \left[V\left(\tau-1,b-1\right)- V\left(\tau-1,b-k\right)\right]
\\ 
& = 
\begin{cases}
\delta - E^{\text{sav}} + \beta^2 \left[V\left(\tau-2,b-1-k\right)- \right. \\ \left. V\left(\tau-2,b-2 k\right)\right], \hspace{0.3cm} \mbox{if } \omega < \omega(\tau-1, b-k); 
\\
\delta - E^{\text{sav}} + \beta \left[\delta - E^{\text{sav}}\right] \\ \hspace{0.3cm} \mbox{if } \omega(\tau-1, b - k) \le \delta < \omega(\tau-1, b-1);
\\
\delta - E^{\text{sav}} + \beta^2 \left[V\left(\tau-2,b-2\right)- \right. \\ \left. V\left(\tau-2,b- k -1\right)\right], \hspace{0.3cm} \mbox{if } \delta \ge \omega(\tau-1, b-1); \\
\end{cases}	
\end{aligned}
\label{18}
\end{equation}
In the first case since $\delta < \omega(\tau-1, b-k)$, according to the equation
\begin{equation}
\begin{aligned}
\delta & < \omega(\tau-1 -\tau^{\prime}, b-k-k\tau^{\prime}) \\
& < \omega(\tau-1-\tau^{\prime}, b-1-k\tau^{\prime}),
\end{aligned} 
\end{equation} 
the difference can be further written as
\begin{equation}
\begin{aligned}
& \delta - E^{\text{sav}} + \beta^2 \left[V\left(\tau-2,b-1-k\right)- V\left(\tau-2,b-2 k\right)\right] \\
& = \delta - E^{\text{sav}} + \beta^2 \left[V\left(\tau-3,b-1-2k\right)- V\left(\tau-3,b-3 k\right)\right] \\
& = \cdots \\
& = \delta - E^{\text{sav}} - \beta^{\tau-1} \left[F(b - (\tau-1)k -1)\right].
\end{aligned}
\end{equation}
Therefore, when $\delta = E^{\text{sav}} + \beta^{\tau-1} \left[F(b-(\tau-1)k-1)\right]$, the first case in \eqref{18} equals $0$. Accordingly, when $k(\tau-1)+2 \le b \le k \tau$, the WI is calculated as:
\begin{equation}
\omega(\tau, b) = E^{\text{sav}} + \beta^{\tau-1} \left[F(b-(\tau-1)k-1)\right]
\end{equation}

\item If $ b \ge k\tau + 1$,
\begin{equation}
\begin{aligned}
V(\tau, b) = \max\left\{\delta+\beta V\left(\tau-1,b-1\right), \right. \\ \left. E^{\text{sav}}+\beta V\left(\tau-1,b-k\right) \right\}. 
\end{aligned}
\end{equation}

Therefore, the difference between actions is
\begin{equation}
\begin{aligned}
g(\tau,b) & = \delta - E^{\text{sav}} + \beta \left[V\left(\tau-1,b-1\right)- V\left(\tau-1,b-k\right)\right]
\\
& = 
\begin{cases}
\delta - E^{\text{sav}} + \beta^2 \left[V\left(\tau-2,b-1-k\right)- \right. \\ \left.  V\left(\tau-2,b-2 k\right)\right], \hspace{0.3cm} \mbox{if } \delta < \omega(\tau-1, b-k); 
\\
\delta - E^{\text{sav}} + \beta \left[\delta - E^{\text{sav}}\right] \\ \hspace{0.3cm} \mbox{if } \omega(\tau-1, b-k) \le \delta < \omega(\tau-1, b-1);
\\
\delta - E^{\text{sav}} + \beta^2 \left[V\left(\tau-2,b-2\right)- \right. \\ \left. V\left(\tau-2,b- k -1\right)\right],  \hspace{0.3cm} \mbox{if } \delta \ge \omega(\tau-1, b-1); \\
\end{cases}	
\end{aligned}
\label{19}
\end{equation}
Similar with the previous case, the difference equals $0$ when $\delta = E^{\text{sav}} + \beta^{\tau-1} \left[F(b-(\tau-1)k+1) + F(b - k\tau)\right].$
Accordingly, when $b \ge k \tau + 1$, the WI is calculated as:
\begin{equation}
\omega(\tau, b) = E^{\text{sav}} + \beta^{\tau-1} \left[F(b-(\tau-1)k-1) + F(b - k\tau)\right]
\end{equation} 

\end{enumerate}
Therefore, the closed-form expression for the WI (17) holds.
\end{proof}

\section{Proof of Theorem 3}

To prove Theorem 3, for any given offloading scheduling policy that violates the STLW rule, we construct an updated policy that meets the STLW rule. Then we need to show that this updated policy can increase the reward, compared with the original one.  
Assume that the $i$-th user has priority over the $j$-th user based on the STLW rule at the $t^{\prime}$-th time slot with the system state ${\bf{S}}_t^{\prime}$. Let $\Gamma \triangleq \max \left\{\tau_{i,t}, \tau_{j,t}\right\} - 1$, assume we have a policy $\mathcal{G} = \left\{{\bf{u}}_{t^{\prime}}, {\bf{u}}_{t^{\prime}+1}, \ldots, {\bf{u}}_{t^{\prime} + \Gamma} \right\}$ violates the STLW rule and selects the $j$-th user instead of the $i$-th user at the $t^{\prime}$-th time slot. Then we construct an updated policy $\tilde{\mathcal{G}} = \left\{{\bf{\tilde{u}}}_{t^{\prime}}, {\bf{\tilde{u}}}_{{t^{\prime}}+1}, \ldots, {\bf{\tilde{u}}}_{t^{\prime} + \Gamma} \right\}$ as follows.
    
	\begin{enumerate}
		\item At the ${t^{\prime}}$-th time slot, $\tilde{\mathcal{G}}$ selects the $i$-th user instead of $j$. That is, ${\bf{\tilde{u}}}_{t^{\prime}}$ is same as ${\bf{u}}_{t^{\prime}}$ except that its $i$-th component is $1$ and the $j$-th component is $0$.
		\item 
		Denote the set of time slots that the policy $\mathcal{G}$ selects the $i$-th user instead of the $j$-th user after the $t^{\prime}$-th time slot by ${\Pi}\left(t\right) \subseteq \left\{{t^{\prime}}+1, \ldots, \min \left\{d_i, d_j\right\}-1\right\}$. 
		\begin{itemize}
		    \item If the set ${\Pi}\left(t\right)$ is empty, let $\tilde{\mathcal{G}}$ take the same actions as the $\mathcal{G}$ in the following time slot, that is,  ${\bf{\tilde{u}}}_q = {\bf{u}}_q$, for $k = t^{\prime}+1, \ldots, t^{\prime}+ \Gamma$.
		    \item If the set ${\Pi}\left(t\right)$ is not empty, denote the minimal time slot of the set by $t_{\min}$. For the time slots $q = t^{\prime}+1, \ldots, t_{\min}-1$, let ${\bf{\tilde{u}}}_q = {\bf{u}}_q$. However, at the $t_{\min}$-th time slot, the new policy $\mathcal{\tilde{G}}$ selects the $j$-th user instead of $i$. That is,  ${\bf{\tilde{u}}}_{t^{\prime}}$ is the same as ${\bf{u}}_{t^{\prime}}$ except that the $j$-th component is $1$ and the $i$-th component is $0$.
		\end{itemize}
	\label{Definition: STLW Interchaning Policy}
	\end{enumerate}

	

    Take a closer look between policies $\mathcal{G}$ and $\mathcal{\tilde{G}}$, after the $t^{\prime}$-th time slot, we can always find a sequence $\left\{{\bf{\bar{S}}}_k \right\}_{q=t^{\prime}+1}^{t^{\prime}+\Gamma}$  in $\mathcal{\tilde{G}}$ as a comparison to the sequence $\left\{{\bf{S}}_k \right\}_{q=t^{\prime}+1}^{t^{\prime}+\Gamma}$ in $\mathcal{G}$, which satisfies the following condition:
	\begin{itemize}
		\item If the set $\Pi(t)$ is empty, then we have $\left\{{\bf{\bar{S}}}_q \right\}_{q=t^{\prime}+1}^{t^{\prime}+\Gamma} = \left\{{\bf{S}}_q \right\}_{q=t^{\prime}+1}^{t^{\prime}+\Gamma}$.
		\item Otherwise, we have 
		\begin{equation}
			\left\{{\bf{\bar{S}}}_q \right\}_{q=t^{\prime}+1}^{t^{\prime}+\Gamma} = \left\{{\bf{\bar{S}}}_{t^{\prime} + 1}, \ldots, {\bf{\bar{S}}}_{t_{\min}}, {\bf{{S}}}_{t_{\min}+1}, \ldots, {\bf{{S}}}_{t_{\prime} + \Gamma}  \right\},
		\end{equation}
	\end{itemize}
	where $t_{\min}$ is the minimal time slot in the set $\Pi(t)$. 
	Since two policies $\mathcal{G}$ and $\tilde{\mathcal{G}}$ always select an equal number of users to perform offloading and will be identical after the $\left(t^{\prime} + \Gamma\right)$-th time slot.  Therefore, to arrive at the result in Theorem 3, we only need to show that 
	\begin{equation}
	\begin{aligned}
	R\left( {{\bf{S}}_t}, {\bf{u}}_t\right) & + \sum_{q = t^{\prime}+1}^{t^{\prime}+\Gamma} R\left( {\bf{S}}_q, {\bf{u}}_q\right) 
	\\ & \le R\left( {{\bf{S}}_t}, {\bf{\tilde{u}}}_t\right) + \sum_{q = t^{\prime}+1}^{t^{\prime}+\Gamma} R\left( {\bf{\tilde{S}}}_q, {\bf{\tilde{u}}}_q\right).
	\end{aligned}
	\label{eq: modified SLSW difference}
	\end{equation}

	To verify \eqref{eq: modified SLSW difference}, we consider the following two cases. 
		
		\begin{enumerate}
			\item When the set ${\Pi}\left(t\right)$ is not empty, for every pair of system state sequence, $\left\{{\bf{S}}_q \right\}_{q=t^{\prime}}^{t^{\prime} + \Gamma}$ and $\left\{{\bf{\bar{S}}}_q \right\}_{q=t^{\prime}}^{t^{\prime} + \Gamma}$, both policies will result in the same result, i.e., the equality holds in \eqref{eq: modified SLSW difference}.		
				
			\item When the set ${\Pi}\left(t\right)$ is empty.  Whenever the policy $\mathcal{G}$ selects the $i$-th user, it must also select the $j$-th user, for $q = t^{\prime}+1, \ldots, \min \left\{d_i, d_j\right\}-1$. 
            Denote the remaining workload of the $i$-th user after its deadline by $\delta_i$ under the policy $\mathcal{G}$. Similarly, $\tilde{\delta}_i$ is the remaining workload under the policy $\mathcal{\tilde{G}}$. Since the $i$-th user has priority over the $j$-th user at system state ${\bf{S}}_{t^{\prime}}$, it implies that $\tilde{\delta}_i = \delta_i -1$ and $\tilde{\delta}_j = \delta_j + 1$ (according to the Definition \ref{Definition: STLW Interchaning Policy}). Therefore, we have the reward difference under two policies calculated as
				\begin{equation}
				\begin{aligned}
				& V_{\tilde{\mathcal{G}}}^{t^{\prime}+\Gamma} - V_{{\mathcal{G}}}^{t^{\prime}+\Gamma}
				\\ 
				& = \alpha \left\{- F(\tilde{\delta}_i) - F(\tilde{\delta}_j) - \left[- F(\delta_j) - F(\delta_i)\right] \right\}
				\\
				& = \alpha \left\{ \left[{\delta}_i^2 - \tilde{\delta}_i^2 +  {\delta}_j^2 - \tilde{\delta}_j^2 \right] \right\}
				\\
				& = 2 \alpha \left(\delta_j - \delta_i\right) - 2 
				\end{aligned}
				\label{eq: reward difference}
				\end{equation}
				
			Note that according to the STLW rule, we have $0 \le \delta_i < \delta_j$, where both $\delta_i$ and $\delta_j$ are integers. Therefore, when $\alpha \ge 1$ (i.e. focus on task completion), this reward difference is always no less than 0, which implies that the constructed policy $\mathcal{\tilde{G}}$ can achieve more rewards than the original policy $\mathcal{G}$.
		\end{enumerate}

\bibliography{refs}
\bibliographystyle{IEEEtran}

\end{document}